\documentclass[trackchanges]{aastex701}

\usepackage{fix-cm}

\usepackage{newtxtext,newtxmath,comment,xcolor,multirow}

\usepackage[T1]{fontenc}
\usepackage{graphicx}	
\usepackage{amsmath}	

\begin{document}

\title{Rotation Measure Analysis of Shocks and Sloshing Fronts in a Galaxy Cluster Merger Simulation}

\author[gname=Jia-Rou, sname=Liou]{J.-R. Liou}
\altaffiliation{}
\affiliation{Department of Physics, National Tsing Hua University, Hsinchu, Taiwan}
\email[show]{alex0106@gapp.nthu.edu.tw (JRL)}  

\author[gname=Alvina Y. L., sname=On]{Alvina Y. L. On} 
\affiliation{Physics Division, National Center for Theoretical Sciences, Taipei, 106319, Taiwan}
\affiliation{Center of Theory and Computation, National Tsing Hua University, Hsinchu 30013, Taiwan}
\affiliation{Mullard Space Science Laboratory, University College London, Holmbury St. Mary, Surrey RH5 6NT, United Kingdom}
\email[show]{alvina.on@ucl.ac.uk (AYLO)}

\author[gname=H.-Y. Karen,sname=Yang]{H.-Y. Karen Yang}
\affiliation{Physics Division, National Center for Theoretical Sciences, Taipei, 106319, Taiwan}
\affiliation{Institute of Astronomy, National Tsing Hua University, Hsinchu, 30013, Taiwan}
\email[show]{hyang@phys.nthu.edu.tw (HYKY)}

\author{J. A. ZuHone}
\affiliation{Harvard-Smithsonian Center for Astrophysics, Cambridge, MA 02138, USA}
\email[show]{john.zuhone@cfa.harvard.edu (JAZ)}

\collaboration{all}{}

\begin{abstract}

Recent observations of the Fornax cluster show depolarization signatures on megaparsec scales,
    which may be associated with
    shocks
    and/or
    sloshing motions
    during cluster merger
    and/or
    in-fall. 
To investigate the possible
    reasons behind 
    the depolarization,
we carry out analytical
    and full polarized radiative transfer (PRT) calculations
    of
    radio point
    sources
    behind 
    a merging galaxy cluster
    simulated using the FLASH code. With uniform background light, 
    we analyzed the rotation measure (RM) morphology near the shock front and the cluster center, where sloshing cold fronts appear. 
For the shock scenario, 
    we find a local RM enhancement by $\sim56\%$ behind the shock front on megaparsec scales, arising from 
   the compression
   of hot gas
   and magnetic field lines. Behind the sloshing cold front,
    the cluster center shows decrement in RM magnitude by $\sim23.3\%$, as a result of 
    the cancellation effect of randomly-oriented magnetic fields induced by sloshing-driven turbulence.
We find that beam depolarization increases behind shock fronts and across sloshing cold fronts, indicating enhanced magnetic field fluctuations across the plane of the sky in both scenarios. By fully accounting for all radiative transfer coefficients in the PRT calculations, 
the uniform background light becomes more depolarized near the cluster center, with the effect growing more pronounced as background intensity decreases. 
This suggests that synchrotron emission and Faraday rotation of the intracluster medium can significantly influence the polarization of background sources. 

\end{abstract}

\keywords{magnetic fields –- polarization -- radiative transfer -- galaxies: clusters: intracluster medium -- methods: data analysis -- radiation mechanism: non-thermal}


\section{Introduction} \label{sec:introduction}

Magnetic fields in the intracluster medium (ICM), though typically dynamically subdominant on large scales, could play a vital role in delaying the onset of hydrodynamic instabilities and stabilizing interfaces of discontinuities such as merger-induced cold fronts.
However, direct observation of large-scale magnetic fields and their structures in the ICM remains challenging, as these fields are weak.
Therefore, indirect methods are often used to infer the properties of the large-scale magnetic fields \citep[see reviews by][]{Carilli2002ARA&A,Ferrari2008ssr, Widrow2002Reviews}.
A common method is the projected rotation measure (RM), which probes the magnetic field strength along the line of sight via the Faraday rotation effect. 
Faraday rotation causes the polarization plane of linearly polarized light to rotate by an angle $\Delta\varphi = \rm{RM} \lambda^2$, where $\lambda$ is the observed wavelength, due to the differing phase velocities of right- and left-handed circularly polarized waves.
The RM depends on the product of the thermal electron number density and the magnetic field strength along the line of sight.

Typically, RM measurements from extragalactic radio sources passing through the intergalactic medium (IGM) yield values of only a few $\rm{rad\ m^{-2}}$, even though they can reach several hundreds near the Galactic plane due to the influence of the denser, magnetized interstellar medium \citep{Simard-Normandin1980ApJ,Carilli2002ARA&A,Hutschenreuter2022A&A}.
In contrast, 
    some radio galaxies like Cygnus A exhibit much steeper RM gradients and significantly higher RM values, while its polarization fraction remains constant. 
This behavior strongly suggests that the enhanced RM observed near central radio sources in galaxy clusters arises from a foreground screen of dense, magnetized intracluster gas acting as a Faraday screen, rather than from thermal gas intermixed with the synchrotron-emitting plasma. If the latter were the case, strong internal Faraday rotation would cause significant depolarization at lower frequencies. This effectively rules out both Galactic contributions and internal Faraday rotation as dominant sources of the observed RM in these systems \citep{Mitton1971MNRAS,Dreher1987ApJ,Clarke2001ApJ}. Therefore, RM of background sources behind galaxy clusters could be an effective probe for magnetic structures in the ICM.

During hierarchical structure formation, the merger and accretion of galaxy clusters generate large-scale shocks that transport energy and propagate through the ICM at speeds exceeding the local sound speed. Owing to the relatively high sound speed in the hot ICM, these shocks are typically weak, with Mach numbers on the order of 1.5–3 \citep{Ryu2003ApJ,Ha2018ApJ}. However, such shocks can significantly affect the cluster environment. They compress the component of the magnetic field parallel to the shock surface while leaving the perpendicular component largely unchanged, thereby amplifying and aligning magnetic fields along their surfaces. Furthermore, the shock itself generates vorticity, which can further cascade into turbulence and tangle magnetic fields in the downstream \citep{Keshet2010ApJ, van2010Sci,Iapichino2012MNRAS,Bruggen2012SSR,Anderson2021PASA}.

In contrast, gas sloshing in clusters arises from large-scale motions of the ICM within the gravitational potential, often triggered by the infall of a subcluster or galaxy. This process produces cold fronts—sharp contact discontinuities in density and temperature—which can drive shear flows that stretch and amplify magnetic fields carried with the gas due to the magnetic field draping effect \citep{Lyutikov2006MNRAS, ZuHone2011ApJ}. The interface of the cold front is susceptible to hydrodynamic instabilities that, at later stages, induce turbulence that distorts and randomizes the field orientation \citep{Schuecker2004A&A}. Such turbulence can lead to depolarization, as fluctuations in field direction cause cancellation of polarized emission along the line of sight \citep{Carilli2002ARA&A,Burn1966MNRAS,Sokoloff1998MNRAS,Murgia2004A&A,Bonafede2010A&A,Anderson2021PASA,Sur2021MNRAS}.

A recent study of the Fornax cluster by \citet{Anderson2021PASA} identified a bow-shaped enhancement in Faraday depth (FD) (Section \ref{subsec:section1}) on the southwest side of the cluster, with a swept-back feature extending northeast, resembling the signature of a shock front. This suggests that dynamical activity in the ICM may have introduced discontinuities in density and reorienting magnetic field lines into a more ordered structure. Additionally, an uneven mass distribution between the main cluster and the southwestern subcluster Fornax A (NGC1316) has been reported \citep{Drinkwater2001ApJ}. A series of merger events, including the infall of NGC1404, are believed to have triggered sloshing and merger-driven shocks on scales exceeding several degrees ($1^{\circ}=360 \rm{kpc}$) \citep{Machacek2005ApJa,SuNGC14042017ApJ,Su2017AAS, Sheardown2018ApJ, Anderson2021PASA}. The latest MeerKAT Fornax Survey \citet{Loi2025AA} further supports these findings, showing an increase in RM scatter beyond 300 kpc from the cluster center, and revealing a previously undetected high-RM stripe extending from north to south-southwest. Together, these observations underline the utility of RM as a powerful diagnostic tool for investigating the magnetic field structure within galaxy clusters.

\citet{Anderson2021PASA} suggests that the enhanced RM corresponds to a shock front, and that a slightly offset contact discontinuity separates a triangular extension of elevated RM values in the northeast. They propose that this entire feature may result from turbulence eddies generated by a northeast–southwest subcluster merger. Curiously, there is a narrow zone where the polarized background source counts is reduced and the RM values are comparatively lower.
This region may contain turbulence that tangles magnetic field lines, resulting in depolarization \citep{Burn1966MNRAS,Sokoloff1998MNRAS,Murgia2004A&A}. 
The asymmetric, swept-back X-ray morphology toward the northeast in Chandra data, and possibly in ROSAT once accounting for smoothing and contamination from point sources, both hint at merger activities \citep{Jones1997ApJ,Scharf2005ApJ,Anderson2021PASA}.
These images also reveal edges consistent with cold fronts at radii of several hundred kiloparsecs
-- features believed to result from gas sloshing triggered by the infall of NGC 1404 \citep{SuNGC14042017ApJ,Su2017AAS}. \citet{Anderson2021PASA} proposes that the region lacking polarized background sources may result from small-scale structures in the ICM depolarizing the emission. These small-scale magnetic structures are often generated by turbulence, which commonly arises between shock fronts and contact discontinuities, or underneath sloshing cold fronts. Motivated by this connection, we aim to use simulations to interpret the observed depolarization and its relation to turbulence-driven magnetic field fluctuations.

The Fornax cluster is a nearby, low-mass, cool-core system located at a distance of $\sim$19 Mpc, centered on the galaxy NGC 1399 in the southern sky at redshift $z=0.00475$. Despite its proximity, no diffuse radio halo has been detected \citep{Birzan2012MNRAS}. X-ray observations provide important insights into the cluster’s core and inner regions, but the currently available data do not reach the cluster outskirts with sufficient depth to probe for shocks on larger scales \citep{Su2017AAS, Loi2025AA}. This limitation is particularly relevant in light of the bow-shaped RM feature identified by \citet{Anderson2021PASA}, which may trace a shock front extending into the outer regions of the ICM. Regardless, two radio jets are seen protruding from opposite ends of the central galaxy NGC1399 at the cluster core, rising up to 10 kiloparsecs \citep{Shurkin2008MNRAS,Davies2013MNRAS,Su2017ApJ,Su2017AAS}. Moreover, further studies show disturbance in the ICM with asymmetric X-ray and metallicity distribution, and the morphology in the FD pattern \citep{Su2017ApJ,Anderson2021PASA} all points to a disturbed system. It is still unclear how the polarized properties of the Fornax cluster is influenced by the complex effects of shocks and cold fronts during the merger activities. The point sources observed in Fornax are primarily background active galactic nuclei (AGN) located behind the cluster field-of-view \citep[e.g.,][]{Machacek2005ApJ,Anderson2021PASA}. These sources, while not physically associated with the cluster, serve as useful probes of the ICM through Faraday rotation and depolarization effects.

In this paper, we focus on analyzing RM maps of shock and sloshing scenarios  
in the ICM, and try to reconcile with
the signatures seen in
the Fornax observations. To quantify these RM effects, 
    we carry out 
    analytical
    and full PRT calculations
    of radio point sources
    behind a merging galaxy cluster
    simulated using the FLASH code \citep{Fryxell2000ApJS}. 
In the analytical approach,
    we assume that 
    Faraday rotation dominates the polarization behavior in the ICM -- as is the case for most Faraday-thin objects -- 
    and that the RM at each location
    is the cumulative Faraday effect of the line-of-sight magnetic field strength
    and thermal electron number density. 
We also compared our analytical results 
    to full PRT calculations 
    that account for 
    absorption, 
    Faraday conversion,
    and the contribution from 
    non-thermal electrons.

The paper is organized as follows. 
In Section \ref{sec:Method}, we introduce the  simulation of a binary cluster merger, in which an infalling subcluster generates shock waves and later induces a gas sloshing event (Section \ref{subsec:FLASH}). We want to look specifically
at how shocks and sloshing reshape the RM and polarization structure. We go beyond RM by solving the full PRT equation, which tracks how polarized light evolves through the ICM, accounting for emission, absorption, and
Faraday rotation (Section \ref{subsec:section1}). We then simplify the model under the assumption of dominant Faraday rotation. We outline the analytical formalisms we apply to compute the polarization and projected X-ray emissivity in Section \ref{subsec:polarization_formulism}. In Section \ref{sec:results}, we present the results of our analysis, focusing on the RM and beam-smoothing effects associated with the shock and sloshing scenarios (Sections \ref{subsec:shock} and \ref{subsec:sloshing}, respectively). We compare polarization signatures derived from the PRT equations across different physical conditions, emphasizing their implications for magnetic field structures and their connection to the observed polarization of background light (Section \ref{subsec:PRT_run}). In Section \ref{sec:discussion}, we discuss the limitations of both the simulation and analytical approaches, and relate our findings to observational data where possible. Finally, in Section \ref{sec:conclusion}, we summarize the main conclusions of this work. 


\section{Method} \label{sec:Method}
\subsection{Galaxy Cluster Merger Simulation using FLASH} \label{subsec:FLASH}
To understand the effects of shocks and sloshing on the distribution of magnetic fields, gas, and in turn the projected polarization signals, we analyze one magneto-hydrodynamic (MHD) binary cluster merger simulation adapted from \cite{ZuHone2011ApJ} using the FLASH code \citep{Dubey2009arXiv,Roediger2012MNRAS}.
The merger simulation is carried out in a box with a length of $L=8$ Mpc, and is comprised of the merger of two clusters on the $x-y$ plane with a 5:1 mass ratio. 
The infalling subcluster is represented by a dark matter (DM) halo devoid of gas. 
The initial separation and impact parameter are 3 Mpc and 0.5 Mpc, respectively, and the subcluster itself is moving in the (-$x$)-direction at $1466 ~\rm km/s$ initially.
While the main cluster has a mass comparable to the Perseus cluster, the properties of shocks and sloshing signatures generated in this simulation are chiefly determined by the mass ratio and impact parameter, 
    thus not sensitive to the cluster masses. 

The adaptive mesh refinement (AMR) method of the FLASH code, which is based on the second order derivative of density,  allows us to have a finer resolution at the region of interest where shocks and sloshing are taking place. 
The gas is set as an ideal fluid with an adiabatic index of $5/3$, and for simplicity, there is no viscosity included in the run. 
The initial profile of the dark matter follows the Navarro–Frenk–White profile (NFW profile) \citep{Navarro1997ApJ}, and their velocity distribution is initialized by the Eddington formula \citep{Eddington1916MNRAS}. 
The gravitational potential is determined by the dark matter, 
    which is represented by a group of collisionless particles, 
    and their dynamics
    are computed by an N-body scheme in the FLASH code. 
Lastly, the magnetic field is initialized as a tangled field with a Kolmogorov power spectrum. The spectrum includes cutoffs at both low and high spatial scales: the lower cutoff corresponds to 500 kpc, while the higher cutoff corresponds to $43 ~\rm kpc$ \citep{ZuHone2013ApJ}. The field strength is normalized to an approximately uniform plasma beta of $\beta = p_{\rm gas} / p_{\rm B} = 100$ \citep{Govoni2010A&A,ZuHone2011ApJ}.

The evolution of the cluster merger process is briefly described as follows. Since the subcluster itself is not equipped with gas, to track its position, we identify the subcluster by the local minimum of the gravitational potential away from the cluster center. Viewed along the $z$ direction, a subcluster comes from the upper right side at $[3, 0.5, 0]$ Mpc towards the lower left side of the simulation box. At time $t=2$ Gyr, the merging event of the subcluster at high speed through the ICM produces a discontinuous density jump across the shock front at about $[-1.1,-0.5,0]$ Mpc, leaving a wake of disturbed gas behind. After the subcluster passes through the cluster center at $t = 2.7$ Gyr, the gas starts to slosh around the gravitationally perturbed DM potential of the main cluster, lagging behind slightly due to the ram pressure it experiences. Since the sloshing gas in the outer layer is hardly mixed with the center because of their density and temperature differences, there are discontinuities in electron density and temperature, forming the so-called ``cold fronts''. 

\subsection{The Polarized Radiative Transfer (PRT) Equations and Rotation Measure (RM)} 
\label{subsec:section1}

In the local frame,
    the PRT equations
    can be written as
\begin{equation}
\frac{d}{d {\rm s}}\begin{bmatrix}
    I_\nu\\
    Q_\nu\\
    U_\nu\\
    V_\nu\\
\end{bmatrix}
=- \begin{bmatrix}
\kappa_\nu & q_\nu & u_\nu & v_\nu \\
q_\nu & \kappa_\nu & f_\nu & -g_\nu \\
u_\nu & -f_\nu & \kappa_\nu & h_\nu \\
v_\nu & g_\nu & -h_\nu & \kappa_\nu \\
\end{bmatrix} \begin{bmatrix}
    I_\nu\\
    Q_\nu\\
    U_\nu\\
    V_\nu\\
\end{bmatrix} + \begin{bmatrix}
    \epsilon_{\,\rm I,\nu} \\
    \epsilon_{\,\rm Q,\nu} \\
    \epsilon_{\,\rm U,\nu} \\
    \epsilon_{\,\rm V,\nu} \\
\end{bmatrix}
\label{eq:PRT}
\end{equation}
\citep[see e.g.][]{Pacholczyk1970Book, Pacholczyk1977Book, Jones1977ApJ},
    where $[I_\nu, Q_\nu, U_\nu, V_\nu]$
    are the Stokes parameters
    at a specific frequency $\nu$
    propagating along
    the distance $s$,
    with the coefficients
    $\kappa_\nu$, $q_\nu$, $u_\nu$, $v_\nu$ 
    for absorption, 
    $f_\nu$ for Faraday rotation,
    $g_\nu$ and $h_\nu$ for Faraday conversion, 
    and $\epsilon_\nu$ for emission
    \citep[see also][for a summary]{YLOn2019MNRAS,Chan2019MNRAS}. 
The scattering term is neglected 
    in this work for
    the diffuse ICM. 
Hereafter, the subscript $\nu$ is omitted for simplicity
    and the specific Stokes parameters are now
    written as 
    $[I, Q, U, V]$.
The polarization angle
is defined in terms of
$\varphi=0.5 \tan^{-1} ({U}/{Q})$, 
    while the degree of linear polarization (DOLP)
    is
    $\Pi_{\rm L} = \left(\sqrt{Q^2+U^2}\right)/ {I}$.

Neglecting absorption,
    emission
    and circular polarization,
    equation~\ref{eq:PRT}
    can be simplified into
\begin{equation}
    \frac{d}{d {\rm s}}
    \begin{bmatrix}
        Q\\
        U\\
    \end{bmatrix} = -\begin{bmatrix}
        0 & f \\
        -f & 0 \\
    \end{bmatrix}\begin{bmatrix}
        Q\\
        U\\
    \end{bmatrix} \,.
    \label{eq:RM_derived}
\end{equation}
In most galaxy clusters,
    the Faraday rotation is mainly contributed
    by the thermal electrons
    \citep{Kotarba2011MNRAS},
    such that
\begin{equation}
    f_{\rm th} 
    = 
    \frac{1}{\pi} 
    ~\left(\frac{e^3}{m^2 c^4}\right)
    ~n_{\rm e,th} 
    ~B_{\parallel} 
    ~\lambda^2,
    \label{eq:Faraday_coefficient}
\end{equation}
where $e$ is the electron charge, $m$ is the mass of the electron, $c$ is the speed of light, $n_{\rm e,th}$ is the number density of the thermal electrons, and $B_{\parallel}$ is the magnetic field strength along the line of sight.

Substituting
    equation~\ref{eq:Faraday_coefficient}
    into \ref{eq:RM_derived} 
    gives
\begin{equation}
    \varphi = \varphi_0+\frac{2\pi e^3}{m^2 {(c\omega)}^2}\int_{\rm source}^{\rm observer} n_{\rm e,th}(\rm s') B_{\parallel}(\rm s')\,\rm ds',
    \label{eq:rotation angle}
\end{equation}

\noindent where $\varphi_0$ is the intrinsic polarization angle of the source. The observed FD is acquired by identifying the peak in the Faraday dispersion function (intrinsic polarized flux as a function of the FD). When there is only a single source along the line of sight, RM is then equivalent to the corresponding FD value at the peak of the Farady dispersion function, and the two terms are interchangeable \citep{Heald2009}. Thus, in this context, the RM is computed by
\begin{align}
    {\rm RM} 
    &= 
    (\varphi -\varphi_0) 
    ~\lambda^{-2}
    \\
    &\approx
    0.812 \int_{\rm source}^{\rm observer} 
    \left(\frac{\rm n_{e,th}(s')}{\rm cm^{-3}}\right)
    \left(\frac{B_{\parallel}(\rm s')}{\mu \rm G}\right)\,
    \left(\frac{\rm ds'}
    {\rm pc}\right) 
    ~{\rm rad~m}^{-2} \,,
    \label{eq:RM}
\end{align}
where we set
$\lambda \approx 0.21$~m
at 1.4 GHz
in our calculations.

To compute polarization signatures, we performed post-processing on simulation snapshots at chosen redshifts. We constructed a pipeline for the PRT analysis by generating a 3D grid ($256^3$ cells) of thermal electron number density, temperature, and magnetic field within the simulation domain. These inputs were used by the PRT code to compute local transfer coefficients. The coupled PRT equations were integrated using a fourth-order Runge-Kutta method, with initial Stokes parameters $[I, Q, U, V]$ specified at the first slice and propagated through the domain to the final plane.

\subsection{Calculation of the Synthetic X-ray and Radio Maps}\label{subsec:polarization_formulism}

To investigate the distribution of gas density and magnetic fields in the ICM during a merger, we construct synthetic X-ray and radio maps. Their corresponding two-dimensional maps are generated by projecting the X-ray and radio emissivities along a given line of sight. For the X-ray component, we calculate the projected emissivity in the energy range 0.5–7 keV using version 13.vr of Cloudy (last described by \cite{Ferland2013RMxAA}), coupled with the astrophysical visualization and analysis toolkit yt \citep[see more details on the astrophysical visualization toolkit from][]{Turk2011ApJS}, assuming gas with a constant $0.3 Z_\odot$.
For the radio emissivity
    we assume that
    the relativistic electrons
    have a power-law energy spectrum
    and
    calculate
    the total radio emissivity
    following
    the approach in
    \cite{Barnes2018MNRAS}.
Here we assume that
    the Lorentz factor of the low energy electron cut-off is 
    $\gamma=10^3$
    and 
    their energy density is one percent of the thermal energy density. 
In terms of radio polarization,
    the complex linear polarization $\textbf{P}$
    can be expressed in terms of the normalized Stokes parameters $Q$ and $U$,
    and hence
    can be written as
\begin{equation}
    \textbf{P} 
    =
    p \exp({i2\varphi)}
    =
    p ( \cos(2 \varphi) + i \sin (2 \varphi) ) \,,
    \label{eq:complex_linear_pol}
\end{equation}
where $p$ is the degree of linear polarization $\Pi_{\rm L}$ of the background light.

In Sections~\ref{subsec:shock} and \ref{subsec:sloshing},
    we assume that 
    the intervening cluster predominantly
    contributes to Faraday rotation only
    and does not emit any radio synchrotron radiation.
For simplicity, we consider 
    a uniform background of radio point sources  
where we assume that every point source has $U = 0$ and $Q = I$,
    such that their initial polarization angle is
    $\varphi_0 = 0$
    and 
    $\Pi_{\rm L} = 1$.
The intervening cluster can be treated
    as a random external Faraday screen
    and the resulting complex linear polarization is then
    \begin{equation}
    \textbf{P} = \textbf{P}_{\rm int} ~\textbf{P}_{\rm ext} 
    \label{eq:linear_complex_polarization}
\end{equation}
    \citep{Sokoloff1998MNRAS},
    where $\textbf{P}_{\rm int}$ 
    is the intrinsic complex polarization
    of the background light source.

    The complex polarization of the external Faraday screen is

\begin{equation}
    \textbf{P}_{\rm ext} = \langle ~\exp{(2i ~{\rm RM} ~\lambda^2)} ~\rangle_{w},
    \label{eq:complex polarization}
\end{equation}
where $\langle ... \rangle$ is the spatial averaging over 
    the telescope beam size $w$.
The spatial averaging of polarization signals
    is similar to the smearing of signals by a telescope with a finite resolution,
    which is also known as beam depolarization.
We assume a Gaussian beam in our calculations.

\section{Results} 
\label{sec:results}

\subsection{Shock Scenario} 
\label{subsec:shock}
We present the RM maps of the merging cluster and examine how the polarization of the uniform background light changes as the merger evolves. For both the shock and sloshing scenarios, in Sections \ref{subsec:shock} and \ref{subsec:sloshing} respectively, we focus on the local region where both processes are taking place and compare the results before and after the events. In order to have a comprehensive picture, we also carry out full PRT calculations in Section \ref{subsec:PRT_run}, including effects such as absorption, emission, Faraday rotation and Faraday conversion. 

\label{shock}

\subsubsection{Enhanced RM magnitude Behind Shock Front}

\begin{figure*}[ht!]
\plotone{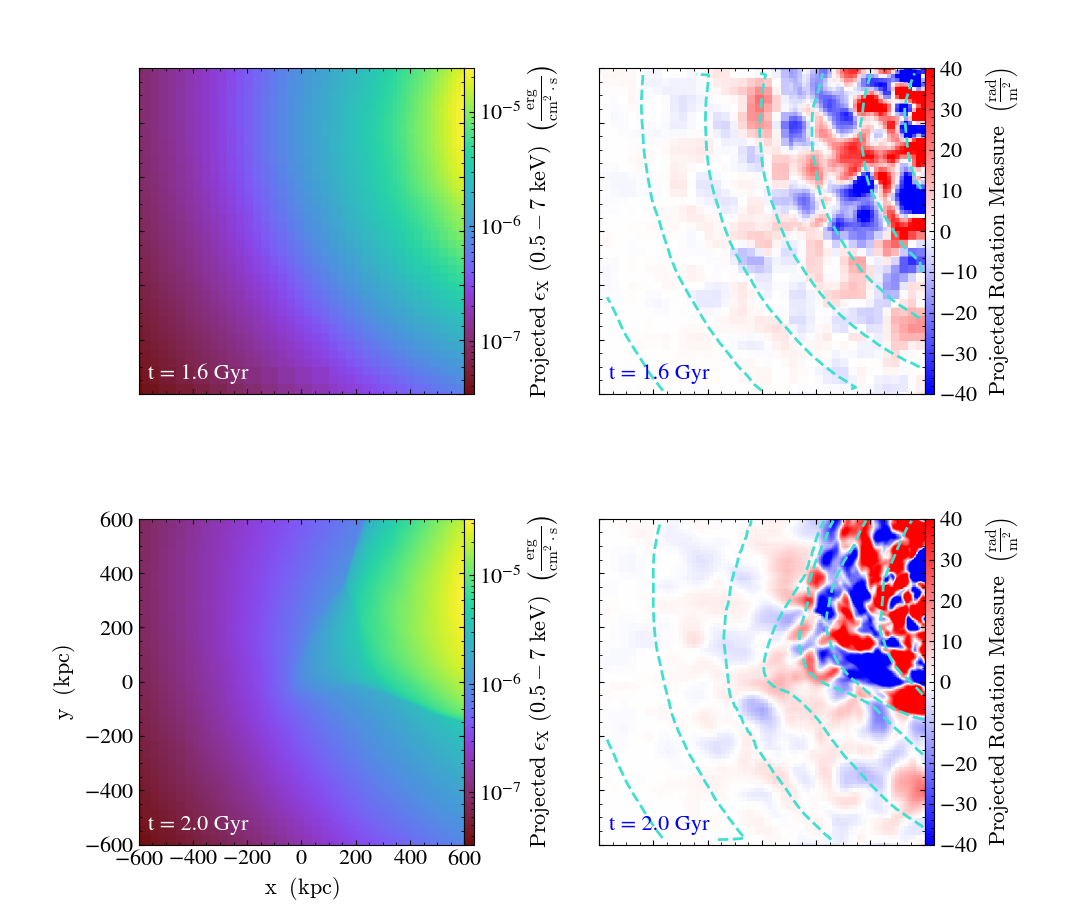}
\caption{Projected X-ray emissivity in energy band $0.5 - 7$ keV (left column) and RM (right column) zoomed-in 
    towards the shock front 
    at 1.6 Gyr (pre-shock; first row)
    and at 2 Gyr (post-shock; second row).
    Contours represent evenly-spaced X-ray surface brightness from -5 to -7.5, with a step size of $d=-0.5$ on a logarithmic scale. We find RM enhancement behind the shock front, while the depolarization effect is not apparent.}
\label{fig:RM_xRay_shock_figure}
\end{figure*}

We compare the RM values centered near the shock front at two different times: 1.6 Gyr, before any significant interaction occurs, and 2 Gyr, during the subcluster's passing. At 2 Gyr, we identify the tip of the shock front at approximately $[-1.1,-0.5,0]$ Mpc  where it appears as a bow-shaped, sharp discontinuity in the projected X-ray emissivity map along the $z$ direction
(see Figure~\ref{fig:RM_xRay_shock_figure}, bottom right panel). 
This feature is observed in the energy band of 0.5–7 keV.

Figure~\ref{fig:RM_xRay_shock_figure} presents the RM maps before and after shock passing overlaid with projected X-ray contours in the right column, while the left displays the simulated projected X-ray emissivity along the $z$ direction. We estimate general variation in RM magnitude in the post-shock region by evaluating the difference in $\mid {\rm RM} \mid$ before and after shock passage, averaging over space. The post-shock region is determined by the pixels enclosed by the X-ray contour marks at the level of $10^{-6} {\rm erg ~{{cm^{-2}}s^{-1}}}$. At 2 Gyr, the magnitude of RM behind the shock front has a local $7.8 ~{\rm rad}\rm ~m^{-2}$ increment on average over the chosen area, which is about a general $56.0\%$ growth after the shock has passed through. This is also associated with larger fluctuations in RM, as its standard deviation increases from $18.0$ to $30.3 ~{\rm rad}~\rm m^{-2}$. The increase in RM magnitude is expected, as both the jump in electron density and the compression of magnetic fields along the shock front enhance the local RM \citep{Lyutikov2006MNRAS,van2010Sci,Iapichino2012MNRAS,Bruggen2012SSR}.

\subsubsection{Enhanced Magnetic Field Strength Behind Shock Front}
\begin{figure*}[ht!]
    
    \includegraphics[width=\columnwidth]{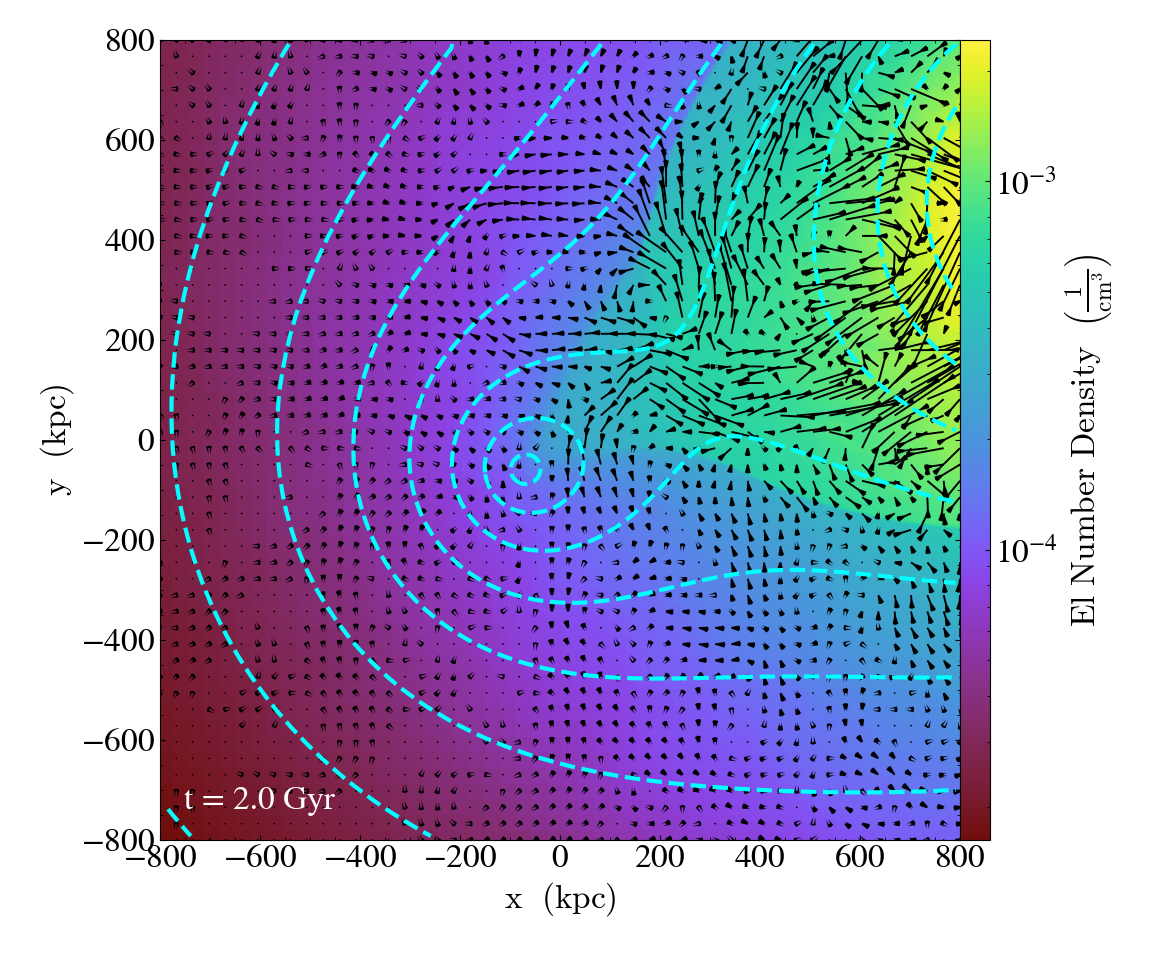}
    \caption{Slice plot of gas density across the shock front, with gravitational potential contours (cyan dashed lines) overlaid with magnetic field vectors (black arrows). The subcluster's central position (dip in the gravitational potential) coincides with the tip of the shock front, indicating the lack of separation between the shock front and the contact discontinuity in our merger simulation.}
    \label{fig:perseus_El_mag}
\end{figure*}

Observations of the FD in the Fornax cluster reveal a depletion of polarized light sources between possible locations of the shock front and the contact discontinuity \citep{Anderson2021PASA}. This depolarized region consists of two distinct parts along the wings of the swept-back patches, curving concavely upward from the center marked by the X-ray contours of the central galaxy NGC 1399. This raises the question of whether the observed depletion results from depolarization, potentially caused by turbulence behind the shock front \citep{Burn1966MNRAS,Murgia2004A&A,Bonafede2010A&A,Ha2018ApJ,Anderson2021PASA}. 
In our simulation, the distinction between the shock front and the contact discontinuity is almost nonexistent. As shown in Figure \ref{fig:perseus_El_mag}, the subcluster that is identified as the point of lowest gravitational potential lies directly atop the shock front, and we observe no clear signs of depolarization.

In Figure~\ref{fig:perseus_El_mag}, the magnetic field strength in the post-shock region is clearly enhanced, as indicated by the longer vectors compared to the ambient medium. While the field orientations appear somewhat disordered—a consequence of the initially tangled field structure and the relatively weak (low-Mach) shock—the overall enhancement is consistent with compression of the field component parallel to the shock front. Over time, this compressed field can evolve into turbulence as vorticity develops behind the shock. The realignment and amplification
of the parallel field component are best visualized in three dimensions, and may not be readily apparent in a 2D slice plot.

\subsubsection{Polarization Fraction with Beam Smoothing}
We examine the effects of beam smoothing on the RM fluctuations across the sky plane before and after the passage of the shock. Before smoothing, we re-sample the AMR grid into a uniform grid with a pixel size of 31.25 kpc. This is because the shock front at 1.6 Gyr lies slightly off-center, and the resolution there is coarse before the shock passage. We therefore re-grid the region around the shock before and after the shock passage to enable a fair comparison.
Gaussian smoothing is then applied using 2D kernels characterized by their standard deviation $\sigma$, which we evolve through $
\sigma= 0, 0.5, 1$ pixel. 

\begin{figure*}[ht!]
    \centering
    \includegraphics[width=\textwidth]{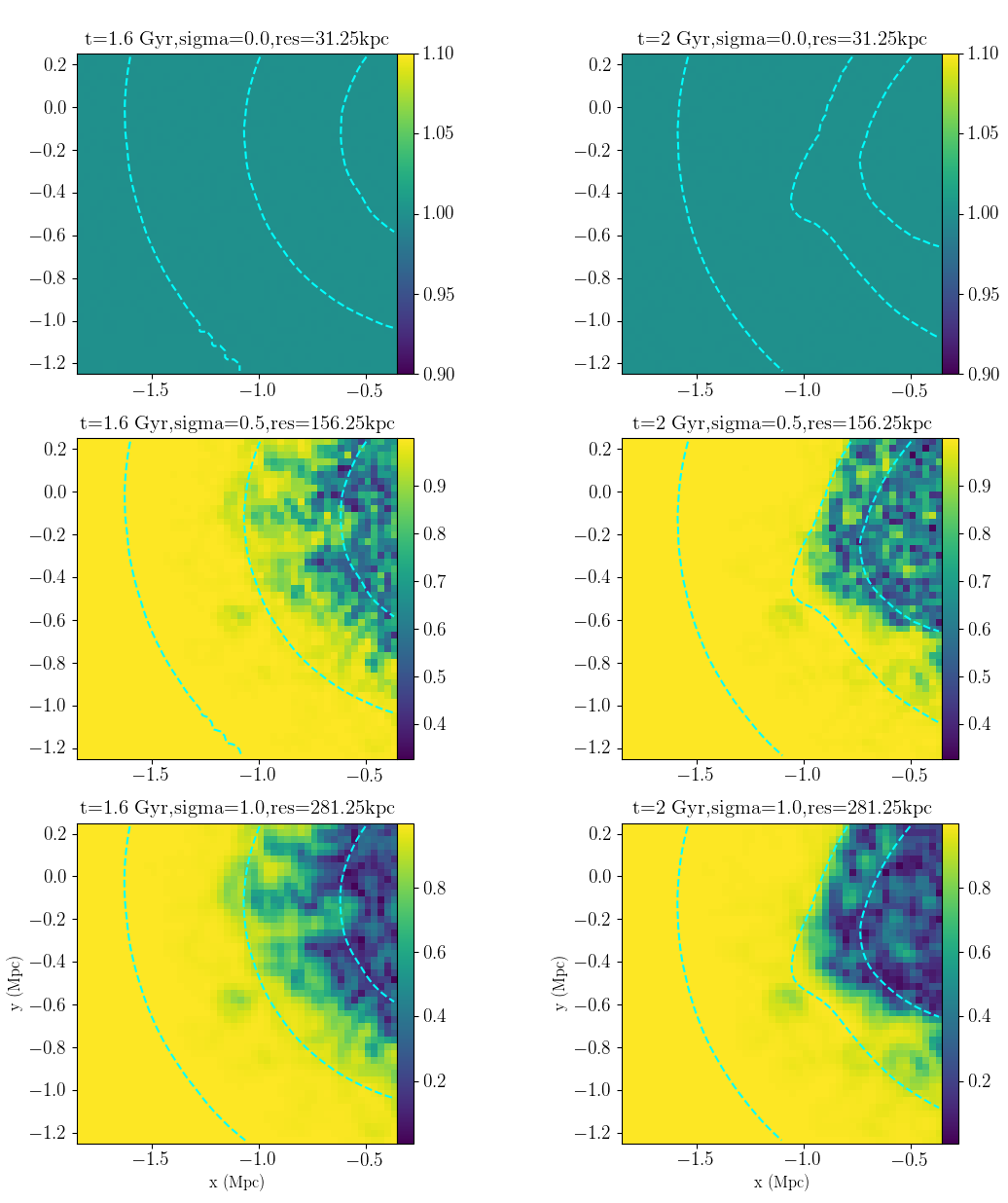}
    \caption{
    The left and right columns show DOLP maps at 1.6 Gyr (before shock passage) and 2 Gyr (after shock passage),
    respectively. The DOLP maps 
    are
    smoothed by sigma (FWHM) of 0 (no smoothing; top row),
    $0.5\times 0.5$ pixels 
    (middle row)
    and 
    $1\times 1$ pixels
    (bottom row). The overlaid contours of X-ray surface brightness in cyan dashed lines represent evenly-spaced values from -5 to -7, with a step size of $d=-1$ on a logarithmic scale. 
    Beam depolarization occurs behind the shock front due to an enhance in magnetic field fluctuations across the plane of the sky, with the effect becoming more pronounced as the beam size increases.}
    \label{fig:DOLP_shock}
\end{figure*}

The resulting DOLP of the shock region subject to different beam smoothing sizes are shown in Figure~\ref{fig:DOLP_shock}. 
In the first row at $\sigma=0$ at 1.6 Gyr and 2 Gyr, 
the smoothed DOLP of the ICM 
is the same as the
initial totally polarized background.
This is because, at $\sigma=0$, there is no beam smoothing, and the DOLP of the ICM is dominated by the 100\% linearly-polarized background.
In the middle and bottom rows, corresponding to smoothing kernels of $\sigma = 0.5$ and $1$, the DOLP centered on the shock front decreases from approximately 0.7 to 0.3. 
This suggests that, as beam resolution worsens, the ability to resolve small-scale structures diminishes. 
The depolarized region 
traces
the shock front 
more clearly
at $2~\rm Gyr$ compared to $1.6~\rm Gyr$.
Notably, both profiles with 
larger
smoothing kernels ($\sigma = 0.5$, $1$) exhibit signs of depolarization even prior to the shock's arrival. This early depolarization is likely caused by fluctuations within the initially tangled magnetic field.  

\subsection{Sloshing Scenario} \label{subsec:sloshing}
\subsubsection{Decreasing RM magnitude after Sloshing}

\begin{figure*}{ht!}
\centering
\includegraphics[width=\textwidth]{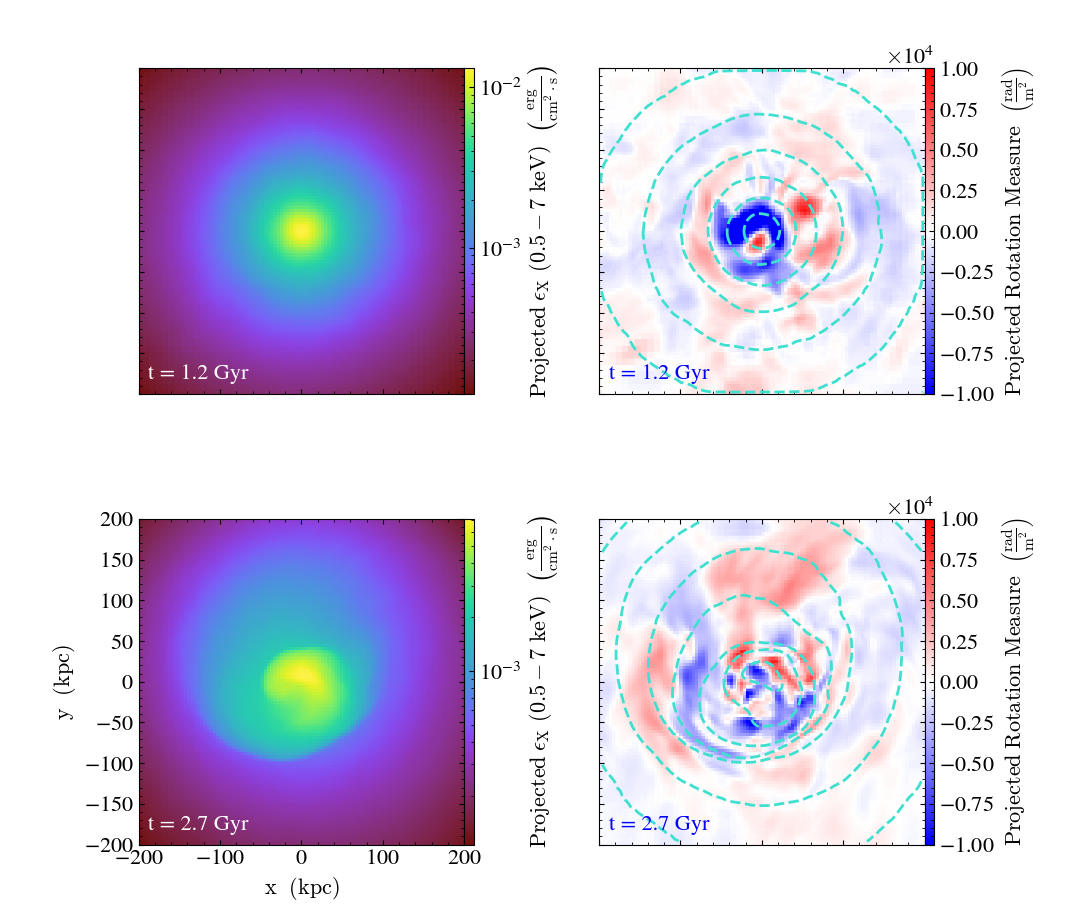}
    \caption{Projected X-ray emissivity in energy band $0.5 - 7$ keV (left column) and RM (right column) zoomed-in 
    towards the sloshing cold front at 1.2 Gyr (at the onset of sloshing; first row)
    and at 2.7 Gyr (after sloshing occurs; second row)
    Contours 
    in the top and bottom panels
    represent evenly-spaced X-ray surface brightness from -2.1 to -3.6 and -2.5 to -3.75, with their respective step sizes being -0.3 and -0.25 on a logarithmic scale. 
    The RM at cold front has in general decreased after sloshing occurs, as a result of
    strong magnetic field entanglement along the line of sight.}
    \label{fig:RM_xRay_slosh_figure}
\end{figure*}

As gas sloshes around the cluster core, 
the sloshing motion may trigger turbulence near the center by Kelvin-Helmholtz instabilities at the interface of the cold front, or the mixing of layers of gas in the presence of a self-gravitational field by Rayleigh-Taylor instabilities. We examine the RM and projected X-ray emissivity maps near the cluster center at t = 1.2 Gyr at the onset of sloshing
(Figure \ref{fig:RM_xRay_slosh_figure} top row);
and at t = 2.7 Gyr during the sloshing motion (Figure \ref{fig:RM_xRay_slosh_figure} bottom row). 
According to \cite{Taylor2002MNRAS} and \cite{Carilli2002ARA&A}, magnitudes of RM in cool-core clusters show an increasing trend with respect to their cooling rates. 
In our simulation,
the highest RM magnitude goes up to more than $10^4~{\rm rad}~{\rm m^{-2}}$, corresponding to the cooling rate of more than a few hundred $\rm M_\odot {\rm yr}^{-1}$, which is 
consistent with the observations of
cool-core clusters at about $100~ \rm{M_\odot} {\rm{yr}^{-1}}$
\citep{Fabian1994ARA&A}. 
In both
X-ray maps at 1.2 Gyr and 2.7 Gyr,  
there is a 
swirling pattern at the center of the cold front, formed by a discontinuous jump in density.
The sloshing region is highlighted by X-ray contours exceeding $10^{-3}{\rm erg}~{{\rm m^{-2}}\rm s^{-1}}$, within which clumps of both positive and negative RM values at 2.7 Gyr appear to fragment into smaller structures. This fragmentation is accompanied by a drop in the average RM magnitude of $-790.9~{\rm rad}~{\rm m}^{-2}$, representing a 23.3\% decrease compared to the the average value at $t=1.2 ~\rm Gyr$.

\subsubsection{Magnetic Field Entanglement Near the Cluster Center}

\begin{figure*}{ht!}
\centering
\includegraphics[width=\textwidth]{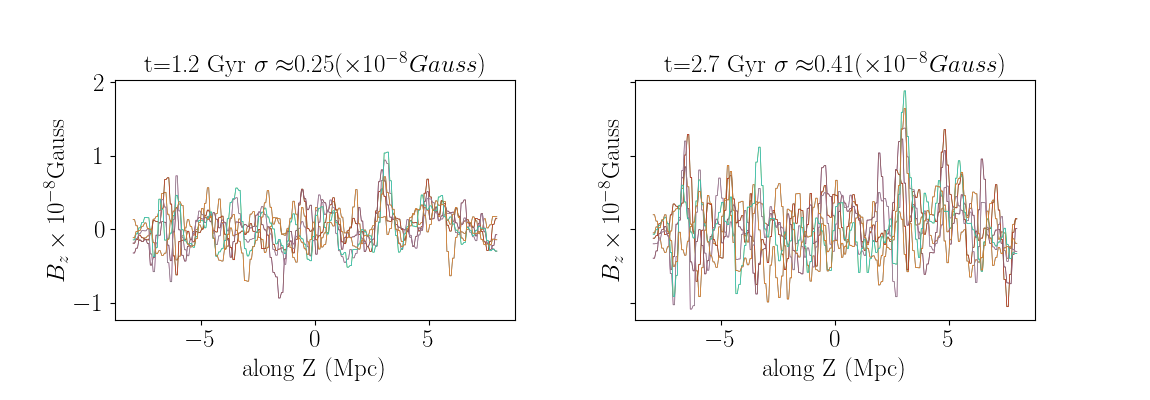}
    \caption{Spatial evolution of the magnetic field’s Z component ($B_z$) along the Z axis, sampled across the X axis from $x = -200~\rm{kpc}$ to $x = 200~\rm{kpc}$, with each color representing different line of sight. This region is experiencing sloshing motions. The left and right panels display $B_z$ profiles at 1.2 Gyr and 2.7 Gyr, respectively. As sloshing progresses, the magnetic field becomes more tangled on smaller scales, as indicated by an increased standard deviation ($\sigma$) of $B_z$. This increased tangling leads to stronger cancellation effects in the RM.}
    \label{fig:B_z_lineplot}
\end{figure*}

The RM magnitude decreases near the cluster center during sloshing at $2.7 ~\rm Gyr$. 
Since the thermal electron number density increases radially towards the center
(see Figure~\ref{fig:RM_xRay_slosh_figure} left column), the large-scale RM decrement 
(see Figure~\ref{fig:RM_xRay_slosh_figure} right column)
must be attributed to the tangled nature of the cluster magnetic field. 
The average RM standard deviation dips near the center from $5122.5$ to $3082.2 ~{\rm rad}~\rm m^{-2}$, implying that the cancellation of magnetic field $z$ components lowers the fluctuation in RM along the $Z$ direction. 
Turbulence and/or gas sloshing can stir the magnetic field, introducing tangled field lines and random field inversions. These effects lead to increased cancellation along the line of sight. 
Figure~\ref{fig:B_z_lineplot}
shows the emergence of more fluctuations in magnetic field $z$ components along the horizontal line cutting across
the cluster center at $\pm 200\, \rm kpc$. 
The local decreases in RM magnitude 
likely arise from these
small-scale magnetic field entanglement, which in turn may contribute to the depolarization of background light passing through the region.

\subsubsection{Lower Polarization Fraction in Cluster Core Revealed by Beam Smoothing}

\begin{figure*}
\centering
	\includegraphics[width=0.74\textwidth] {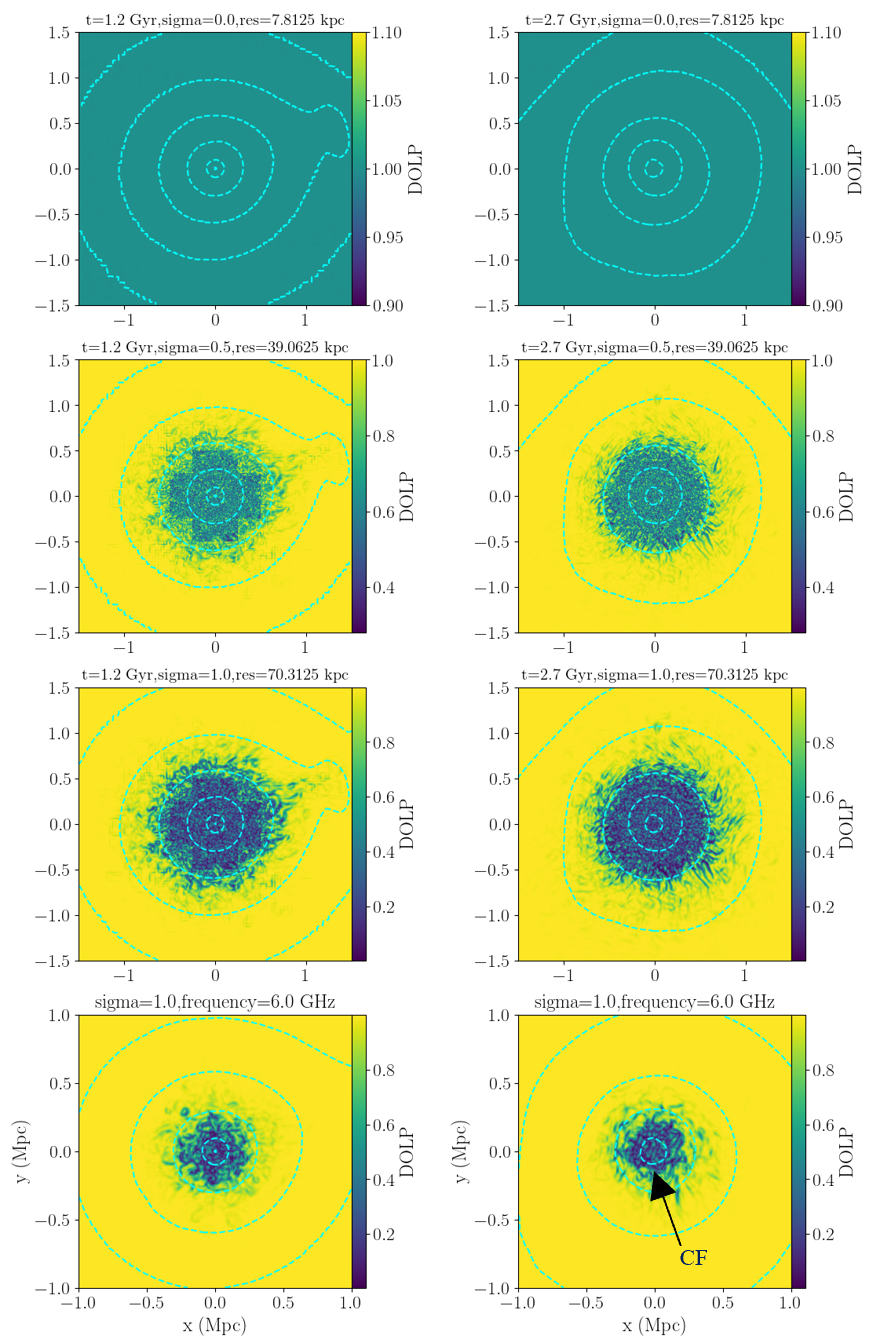}
    \caption{The first three rows show the DOLP maps smoothed by sigma (FWHM) values of $0$ (no smoothing), $0.5\times 0.5$, and $1\times 1$ pixels, observed at $1.4\, \rm Gyr$. The contours represent X-ray surface brightness starting at -3 at the center, decreasing by 1 with each step outwards on a logarithmic scale. The left and right columns show profiles at 1.2 Gyr (the onset of sloshing) and 2.7 Gyr (after sloshing occurs), respectively. The last row highlights depolarized regions at 6 GHz, smoothed with $1\times 1$ pixels and zoomed in on the $500~{\rm kpc}$ central region.
    Beam depolarization occurs at the sloshing center due to enhanced magnetic field fluctuations, becoming more pronounced as beam size increases from rows 1 to 3.
    The depolarization area is similar
    between 1.2 Gyr and 2.7 Gyr, due to the initially tangled magnetic field. Comparing between rows 3 and 4,
    the central region
    is more depolarized 
    at 1.4 GHz than at 6 GHz,
    due to Faraday depolarization
    being more severe
    at lower frequencies. 
    The compressed cold front (CF)
    is visible after sloshing 
    at 6 GHz. }
    \label{fig:DOLP_slosh}
\end{figure*}

In Figure~\ref{fig:DOLP_slosh}, 
    we examine the effects of beam depolarization on the polarization fraction across the sky plane before and after the onset of sloshing at 1.2 Gyr and 2.7 Gyr, respectively. The polarization fraction is calculated analytically using  
    equation~\ref{eq:complex polarization}, where the AMR grid is re-sampled into a uniform grid with a pixel size of 7.81 kpc. We apply a Gaussian smoothing with standard deviation $\sigma$ ranging from $0, 0.5$ to $1 ~{\rm pixel}$.

The smoothed polarization fraction maps zoomed-in at 
the cluster center 
before and after 
the onset of sloshing
are shown in left and right columns of Figure~\ref{fig:DOLP_slosh},
respectively.
The first row corresponds to
$\sigma=0$ 
and shows totally polarized background light with no signals being smeared. 
The second and third rows correspond to 
$\sigma=0.5$ 
and $\sigma=1$, respectively. We notice that the beam-smoothed profile of the sloshing core shows a large, boxy depolarized region, which is likely an artifact of the lower effective resolution at that epoch and sparse sampling in the spherical grid. However, this does not significantly influence the key results described below.
As the beam size increases, 
the level of polarization decreases,
    with the minimum value of DOLP dropping from $0.7$ to about $0.3$.
The fourth row is also smoothed at $\sigma=1$, 
    similar to the third row, 
    but observed at a higher frequency of 6 GHz.
As expected, 
    Faraday depolarization is less severe at a higher frequency
    \citep[see e.g.][]{Sokoloff1998MNRAS,Sur2021MNRAS},
    so the depolarization region
    extends over a smaller length scale at 6 GHz (fourth row) than at 1.4 GHz (third row). 
 By reducing the beam depolarization effect at higher frequency, the depolarization shaped by the sloshing event brings forth a distinctive shape of a compressed cold front.

\begin{figure}
	\includegraphics[width=\columnwidth]{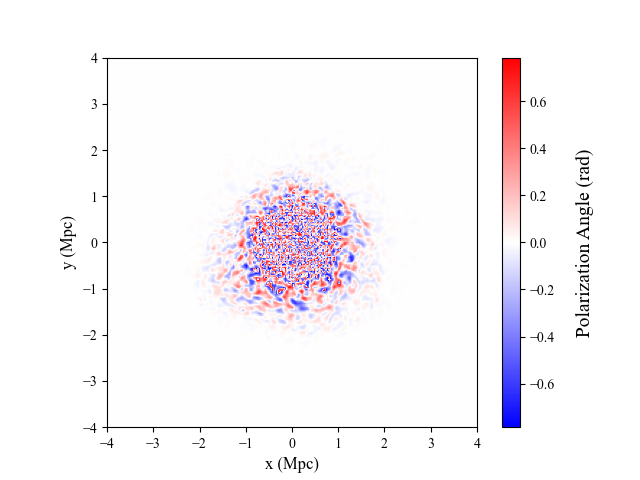}
    \caption{Semi-analytical map of the polarization angles encompassing the whole cluster at 2.7 Gyr. The center exhibits a criss-cross pattern of angles flipping between positive and negative values, resulting from the high RM in that region.
     }
    \label{fig:analytical_PA_t=2.7Gyr_N=256}
\end{figure}

The lower polarization fraction at the cluster center 
before sloshing
arises mainly
from the initial fluctuations in the tangled 
magnetic fields
(c.f. Figures~\ref{fig:RM_xRay_slosh_figure} and \ref{fig:DOLP_slosh}).
Figure~\ref{fig:analytical_PA_t=2.7Gyr_N=256} shows the polarization angles (PA) at $2.7 ~\rm Gyr$, which are calculated analytically 
    using equations~\ref{eq:rotation angle},
    \ref{eq:RM}
    and
    \ref{eq:complex_linear_pol}.
The angles flip randomly between positive and negative values,
    due to the steep RM gradients near the cluster center, where the RM magnitude can range from several hundred to several thousand ${\rm rad}~{\rm m}^{-2}$ across a few pixels. 
These gradients lead to significant variations in the FD along different lines of sight, which, in turn, results in a lower observed DOLP
    in Figure~\ref{fig:DOLP_slosh}.

\begin{figure*}
\centering
	\includegraphics[width=\textwidth]{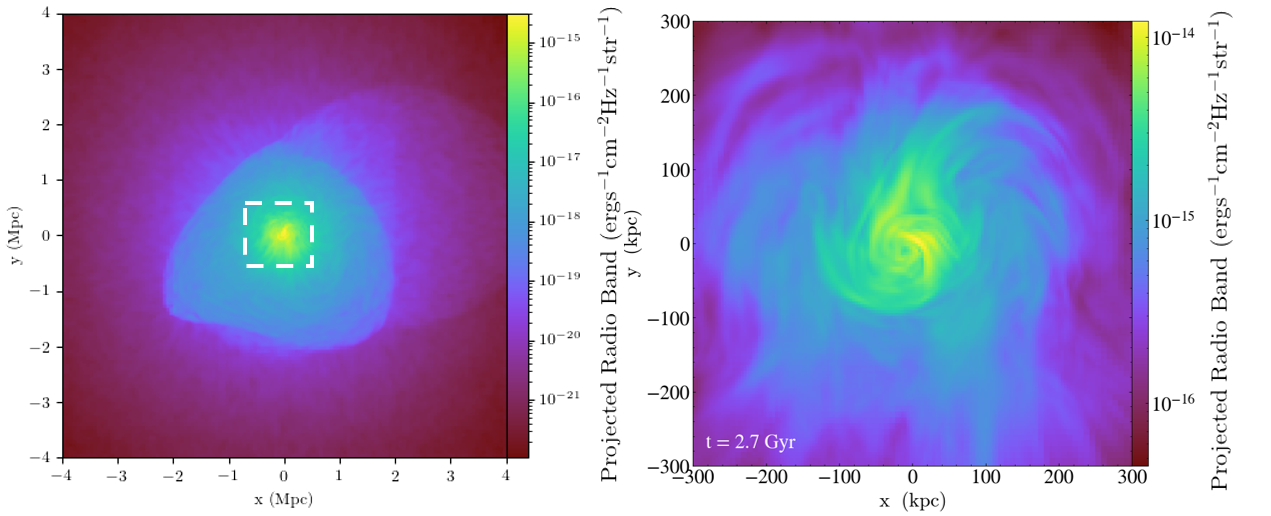}
    \caption{The left panel shows the analytical 1.4 GHz radio synchrotron sky
    map at 2.7 Gyr. 
    The right panel shows the synchrotron emission zoomed in on the cluster center (white dashed box), 
    showing the sloshing pattern which traces the structures of the transverse component of the magnetic field.}
    \label{fig:analytical_synchrotron_sloshing_edit4}
\end{figure*}

\subsection{Comparisons with a full polarized radiative transfer treatment}
\label{subsec:PRT_run}

In Sections \ref{subsec:shock} and
    \ref{subsec:sloshing},
    we calculated 
    the RM 
    and PA maps 
    analytically,
    assuming that
    the cluster
    is optically thin
    and 
    only 
    undergoes
    Faraday rotation.
    These are reasonable assumptions in our case,
    as
    a back-of-envelope comparison 
    of the normalized transfer coefficients
    near the cluster center
    gives
    $f'>>h'>>1>>\epsilon_{Q}'>q'>>\epsilon_{v}'>v'$ \citep[see equation 2 in][for a summary]{Jones1977ApJ}.
    The Faraday rotation coefficient
    $f'$ exceeds the convertibility coefficient $h'$ by approximately an order of $10^7$, and is about  $10^{11}$ larger than the rest of the coefficients.

In reality, 
    many merging clusters
    are complex systems,
    with a fraction of them
    hosting diffuse radio emission
    in the form of halos
    and/or 
    relics.
    There can also be 
    intervening synchrotron sources, 
    such as radio galaxies
    embedded within the medium,
    which
    can absorb, 
    emit 
    and even Faraday convert 
    the linearly-polarized light.
    In this section,
    we generalize our analyses to a broader range of conditions and carry out
    full PRT
    calculations
    of our merging cluster simulation,
    assuming a uniform background 
    of radio point sources
    and compare
    the results 
    to our analytical treatment.

\subsubsection{Consistent Synchrotron Emission and Polarization Angles with Analytical Treatment}

\begin{figure*}
\centering
	\includegraphics[width=\textwidth]{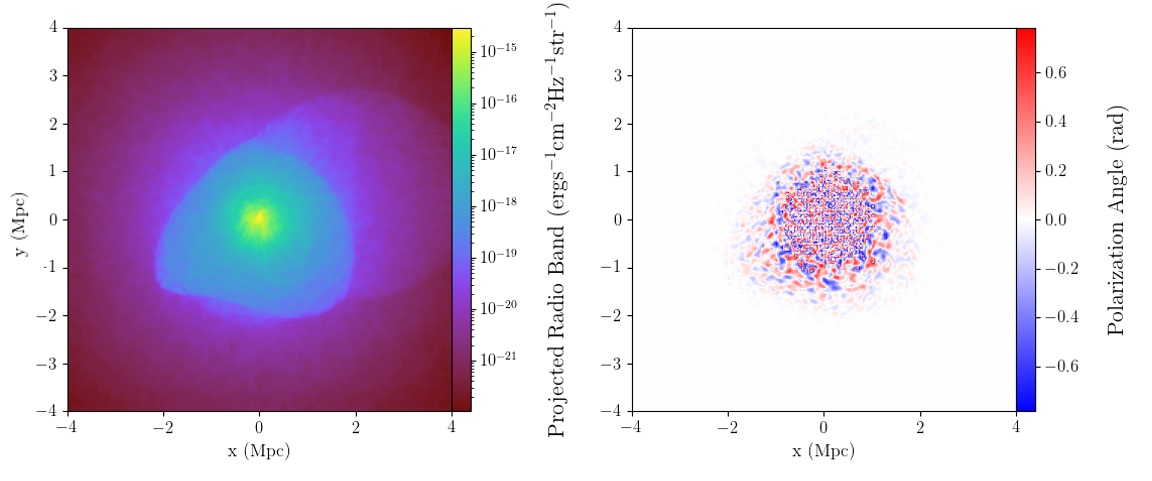}
    \caption{
    The total synchrotron intensity map (left panel) and polarization angle map (right) of the cluster at 2.7 Gyr generated
    using
    PRT calculations,
    assuming that 
    only
    emission coefficient $\epsilon_I$ 
    and the Faraday rotation coefficient $f$ 
    are turned on.
    The analytically-implemented and PRT-computed synchrotron and polarization angle signatures (c.f. Figure~\ref{fig:analytical_PA_t=2.7Gyr_N=256}) are consistent.
     }
    \label{fig:PRT_run_PA_synchrotron}
\end{figure*}

We manually implement the analytical form of synchrotron emissivity (Section \ref{subsec:polarization_formulism}). Figure~\ref{fig:analytical_synchrotron_sloshing_edit4}
    (left)
    shows the analytical 1.4 GHz 
    total radio 
    synchrotron sky map at 2.7 Gyr.
By zooming in on the cluster center (right),
    we see 
    spiral features from the sloshing motion,
    where the discontinuity at the underside of the spiral forms a cold front. 
For a comparison,
    we carried out a PRT calculation
    of the same system,
    without any background sources.
Assuming that only emission
    and Faraday rotation
    are present,
    we calculated the total synchrotron intensity map (left) and 
    polarization angle map (right)
    of the cluster
    at 2.7 Gyr in 
    Figure \ref{fig:PRT_run_PA_synchrotron}.
The analytical 
    and PRT-computed
    synchrotron intensity
    maps
    (c.f. the left panels of Figures~\ref{fig:analytical_synchrotron_sloshing_edit4}
    and \ref{fig:PRT_run_PA_synchrotron})
    and polarization angle maps
    (c.f. 
    Figure~\ref{fig:analytical_PA_t=2.7Gyr_N=256}
    and the right panel of Figure~\ref{fig:PRT_run_PA_synchrotron})
    are consistent
    in terms of
    morphology
    and order of magnitude.

\subsubsection{Polarization Maps near the Sloshing Center with Varying Background Intensity or Varying Background DOLP} 
\label{dolp with intensity}

\begin{table*}
    \centering
    \begin{tabular}{cccccccccclll}
    \hline
          \multicolumn{1}{c}{run}    
         &  \multicolumn{4}{c}{background source}
         &  \multicolumn{5}{c}{transfer coefficients}
         & \multicolumn{3}{c}{emission coefficients}
         \\
         &  \(I_0\)
         &  \(Q_0\)
         &  \(U_0\)
         &  \(V_0\)
         &  \(f\)
         &  \(h\)
         &  \(k\)
         &  \(q\)
         &  \(v\)
         & \(\epsilon_{\rm I}\)
         & \(\epsilon_{\rm Q}\)
         &\(\epsilon_{\rm V}\)\\
         \hline
         f1e0&\multirow{3}{*}{$10^{-(13-15)}$} & \multirow{3}{*}{$10^{-(13-15)}$} & 0&  0&  on&  off&  off&  off&  off& off& off&off\\
         f1e1&  && 0&  0&  on&  off&  off&  off&  off& on& on&on\\
         all1e1&  && 0&  0&  on&  on&  on&  on&  on& on& on&on\\
         \hline
    \end{tabular}
    \caption{
    Set-up of the
    transfer
    and
    emission 
    coefficients for
    thermal and  
    non-thermal
    electrons,
    and the corresponding
    background source
    Stokes parameters used in the polarized radiative transfer (PRT) calculations. 
    We vary the intensity $I_0$ from $10^{-13}$, $10^{-14}, 10^{-15}$ $\rm{erg}\, {s}^{-1} {cm}^{-2} {Hz}^{-1}  {str}^{-1}$ to simulate the different intensities
    in background light, while keeping them highly
    linearly polarized
    at 100\% with $I_0=Q_0$. For run f1e0,
    only the Faraday rotation coefficient
    is switched on,
    whereas
    all of the coefficients are turned on in
    run all1e1.
    In run f1e1, only Faraday rotation
    and emissivities $\epsilon_{\rm I, Q, V}$ are turned on.}
    \label{tab:my_label}
\end{table*}

We calculate
    1.4 GHz 
    mock radio observations
    of 
    the merging cluster simulation
    at 2.7 Gyr
    towards
    a uniform distribution
    of background point sources. 
We vary the intensity of the background point sources 
    $I_0$ from $10^{-13}$, $10^{-14}, 10^{-15}$ $\rm{erg}\, {s}^{-1} {cm}^{-2} {Hz}^{-1}  {str}^{-1}$,
    which is about $100, 10$ and 1 times the maximum synchrotron emission
    from the cluster itself. 
The background point sources are
    assumed to be 
    totally
    linearly polarized
    at DOLP = 100\%,
    with $I_0=Q_0$. 
    We set $U_0 = 0$,
    such that
    the initial polarization angle
    is fixed at $\varphi_0 = 0$ radian. 
The initial conditions
    of the uniform polarized background sources 
    make it 
    more transparent for us to
    keep track
    of any depolarization signatures
    due to the intervening 
    ICM. 

\begin{figure*}
\centering
	\includegraphics[width=\textwidth]{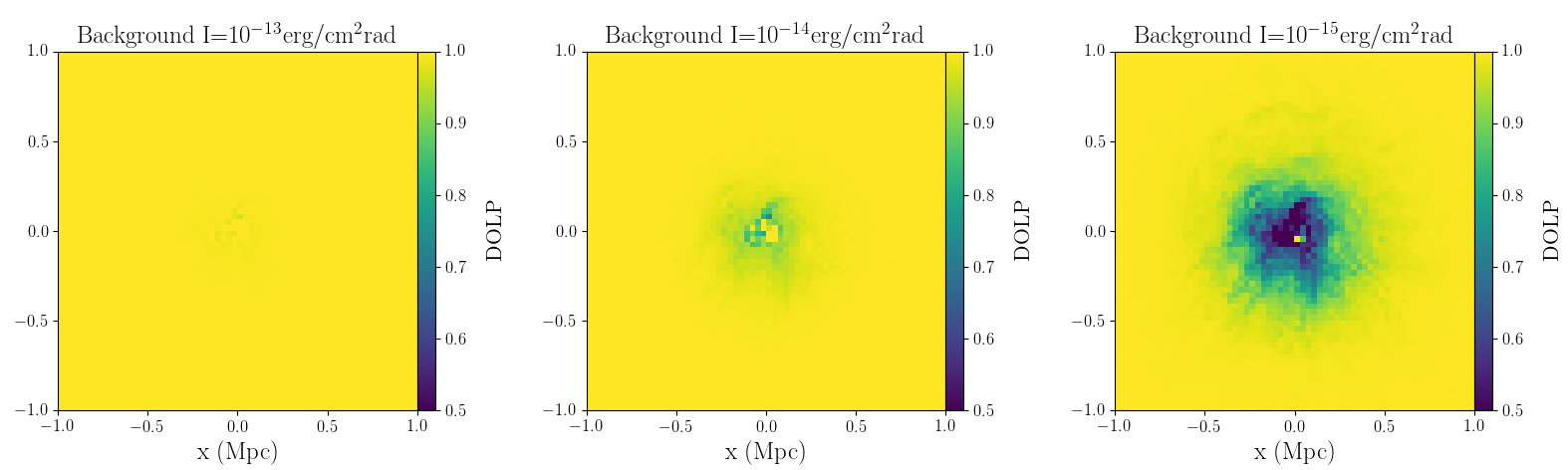}
    \caption{DOLP maps solved by the PRT code with all coefficients turned on, focusing on the sloshing center. From left to right, the initial background intensity $I$ is uniform at
    $10^{-13}, 10^{-14}$, and $10^{-15}\ \rm{erg}\, s^{-1} Hz^{-1}~\rm{cm}^{-2}\ \rm{str}^{-1}$, respectively. We find that uniform background light is more depolarized near the cluster center, with this effect becoming more pronounced as background intensity decreases.
    There are a few pixels near the center where their values are unusually high (the yellow pixel located slightly offset in a sea of green pixels), which could be due to numerical errors. 
    }
    \label{fig:PRT_change_intensity}
\end{figure*}

In Table~\ref{tab:my_label},
we consider three cases,
    where 
    only Faraday rotation occurs
    (f1e0),
    both Faraday rotation
    and emission
    are present
    (f1e1),
    and
    all emission 
    and transfer effects
    (including absorption
    and Faraday conversion)
    take place
    (all1e1).
The DOLP maps are similar across cases
    f1e1 and all1e1,
    indicating that 
    absorption
    and Faraday conversion
    are insignificant
    in our system. 
Figure~\ref{fig:PRT_change_intensity} shows the 
DOLP maps for case all1e1,
    as the background intensity
    lowers
    from $10^{-13}$ (left), 
    $10^{-14}$ (middle), 
    to $10^{-15}$ $\rm{erg}\, {s}^{-1} {cm}^{-2} {Hz}^{-1}  {str}^{-1}$ (right).
The depolarization effect becomes more pronounced as the background intensity decreases
    to a similar level
    as the cluster emission.
This is consistent
    with the depolarization conditions
    identified in
    \cite{On2025PASA},
    where the dim sources 
    are depolarized
    by different amounts, 
    depending on the location of the sources
    and the magneto-ionic properties of the cluster.

\begin{figure*}
\centering
	\includegraphics[width=\textwidth]{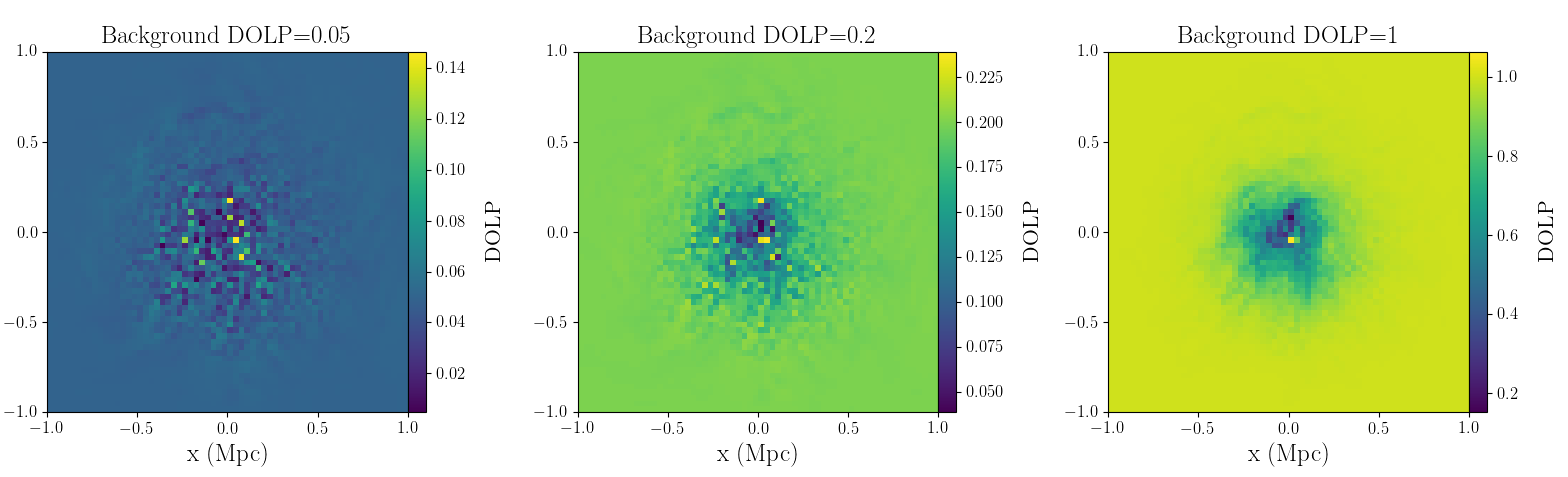}
    \caption{DOLP maps solved by the PRT code with all coefficients turned on, focusing on the sloshing center. The initial background intensity is set to $10^{-15} \rm{erg}\, {s}^{-1} {Hz}^{-1} {cm}^{-2} {str}^{-1}$,
    while the DOLP varies from
    5\%, 20\% to 100\% from left to right.
    We find that uniform background light is sometimes enhanced near the cluster center, with this effect becoming more pronounced as background DOLP decreases.
    }
    \label{fig:PRT_change_DOLP}
\end{figure*}

To quantify how the observed polarization fraction changes
    with initial source DOLP,
    we set a uniform background
    total intensity of 
    $I_0 = 10^{-15} \rm{erg}\, {s}^{-1} {Hz}^{-1} {cm}^{-2} {str}^{-1} $,
    while varying the initial DOLP 
    from 5\%, 20\% to 100\%.
Figure~\ref{fig:PRT_change_DOLP} shows the resulting polarization maps zoomed in on the cluster center. As the DOLP of the background sources increases (from left to right), most cluster sources exhibit a corresponding rise in polarization fraction. A notable depolarization, emerges near a radius of approximately $500~\rm{kpc}$.
In the left and middle panels, however, several pixels stand out with noticeably higher polarization values compared to their local surroundings. 
These pixels do not undergo depolarization;
    instead, their observed DOLP is
    larger than
    their initial DOLP.
This enhancement is physical,
    and it occurs when the polarization fraction of the ICM
    exceeds that of the background light,
    mainly due to
    larger electron number densities and
    stronger, more aligned magnetic fields 
    near the cluster center
    \citep[see also][for polarization enhancement conditions]{On2025PASA}.
In general,
    the number of enhanced pixels decreases as the background DOLP increases.

\section{Discussion} 
\label{sec:discussion}

\subsection{Comparison between the shock and sloshing features in our simulations}

Both shock and sloshing 
are able to cause fluctuations in the magnetic field lines. 
From the beam smoothed 
DOLP maps in Figure \ref{fig:DOLP_shock} and Figure \ref{fig:DOLP_slosh}, both occurrences
lead to beam depolarization behind the shock/cold front, respectively. 
The overall DOLP follows a general trend as it decreases from roughly 0.7 to 0.3 with a larger beam size.
The extra frequency observed at $6 ~{\rm GHz}$ reveals a more distinct shape of the sloshing cold front
    because the fluctuations in FD diminish with higher frequency (shorter wavelength).

The morphologies and the change in the RM magnitude
    due to shock
    and sloshing
    are quite different, however.
There is an 
asymmetric, bow-shaped enhancement 
    of approximately
    $56.0 \%$
    in the selected region for the RM 
    behind the shock front
    in our simulation, 
    due to the 
    compression
    in gas
    and magnetic field lines
    after shock passing (Figure \ref{fig:RM_xRay_shock_figure}).
On the other hand,
    the sloshing motions
    generate a spiral pattern in
    the RM 
    around
    the cluster center,
    and have generally
    lower magnitude of RM
    within the inner radius of $100 ~{\rm kpc}$ .
There is a $23.3\%$ decrease with respect to the initial RM magnitude (Figure \ref{fig:RM_xRay_slosh_figure}).

\subsection{Comparison between our simulation results and observations} 
\label{sim-obser}

The morphologies in the X-ray and RM maps near the shock front in our simulation
    (Figure~\ref{fig:RM_xRay_shock_figure}) 
    closely resemble
    some of the features seen in
    observations
    \citep{Anderson2021PASA}.
The X-ray morphology is disturbed
    in 
    Figure~\ref{fig:RM_xRay_shock_figure} (left column),
    clearly tracing
    the merger axis
    in the NE - SW direction
    as the subcluster falls 
    into the main cluster,
    with X-ray emission
    brightening
    behind the wake.
The projected RMs 
    in our simulation
    are also visibly higher
    in the post-shock region
    and extend
    over
    Mpc scales,
    which are similar to 
    the Fornax observations. 
On the other hand,    
    the RM maps
    in Figure~\ref{fig:RM_xRay_slosh_figure} (right column)
    show that sloshing motions 
    can randomize magnetic field lines,
    causing 
    the RM to change signs on small scales.
It is therefore possible for     
    turbulence 
    to result in
    the isotropic
    distribution
    of RMs seen behind the wake 
    in the Fornax cluster.

There are a few differences between
    our simulations
    and observations.
In our simulation,
    there is no distinct separation between the contact discontinuity (the edge of the subcluster) and the shock front
    (see Figure~\ref{fig:perseus_El_mag}). 
This means that we cannot 
    assess whether there is depolarization or a lack of polarized sources
    between the area
    enclosed 
    by the contact discontinuity
    and the bow shock,
    as proposed in
    \cite{Anderson2021PASA}. 
In addition, 
    our shock has a smaller 
    opening angle
    at $\sim 30^{\circ}$
    with 
    a larger Mach number 
    at $M_{\rm s} \approx 2$  
    than in the Fornax observation. 
Having a stronger shock in our simulation may lead to a larger RM enhancement 
in the post-shock region.

Our simulation does not include viscosity, and it is therefore more susceptible to hydrodynamic instabilities. 
If viscosity is taken into account,
it would suppress turbulence in the ICM, 
    leading to less fluctuations in the RMs.
Moreover,
    viscosity is also an important factor to stabilize cold fronts. As such,
    it is possible to 
    delay the onset of
    turbulence 
    near the sloshing center,
    affecting the level at which the RM magnitude changes in our simulation.

\section{Conclusions}
\label{sec:conclusion}

We carried out analytical
    and full PRT calculations
    of
    radio point
    sources
    behind 
    a merging galaxy cluster
    from the FLASH 
    simulations    
    to investigate
    the reasons
    behind depolarization
    seen
    in the recent Fornax cluster observations.
Our main results are summarized below.

\begin{enumerate}
    \item The projected RMs 
    in our post-shock region
    have a similar morphology
    to the Fornax observations.
    The RMs
    are visibly higher. There is an $\approx 56\%$ increase in RM magnitude in the region
    behind the shock front, which extends
    over
    Mpc scales.
    The enhancement along the shock front likely arises from 
   the compression
   of hot gas
   and magnetic field lines
   in our simulation.
The X-ray contours
    in our simulation 
    follow
    a swept-back morphology
    along the cluster merger axis,
    which is comparable
    to the X-ray observations
    of the Fornax cluster.

    \item The sloshing core within a radius of a few hundred ${\rm kpc}$ shows a decrease in RM magnitude by $\approx 23.3\%$ in the affected region, due to the tangled magnetic field induced by sloshing-driven turbulence. 

    \item The smoothed DOLP of shock front and sloshing center  reveal small-scale, unresolved structures depolarizing the background sources. 
    The beam depolarization region extends around the shock front and the cold front.

    \item 
    By incorporating all relevant radiative transfer coefficients into the PRT calculations, we find that the DOLP of the ICM exhibits more pronounced depolarization features near the cluster center as the background source intensity decreases.  
    For a constant background intensity, the number of pixels 
    with enhanced polarization increases with diminishing initial background DOLP. These findings underscore how synchrotron emission and Faraday rotation within the intervening ICM significantly modulate the observed polarization of background radio sources.

\end{enumerate}

While there is no clear indication 
    from our simulations
    as to why there is a lack of 
    polarized radio sources in 
    the Fornax cluster field,
our work demonstrates
    that deep, wide-field RM
    and polarization grids
    are valuable means
    to better understand
    the magneto-ionic properties 
    of the ICM.

\section*{Acknowledgements}

JRL and HYKY acknowledge support from National
Science and Technology Council (NSTC) of Taiwan
(NSTC 112-2628-M-007-003-
MY3; NSTC 114-2112-M-007-032-MY3). HYKY acknowledges support from
Yushan Scholar Program of the Ministry of Education
(MoE) of Taiwan (MOE-108-
YSFMS-0002-003-P1). 
AYLO is supported by
    the NSTC of Taiwan 
    through the grants
    111-2124-M-002-013-,
    112-2124-M-002-003-
    and
    113-2124-M-002-003-,
    the MoE
    (Higher Education Sprout Project
        NTU-113L104022-1)
    and
    the National Center for Theoretical Sciences of Taiwan.
This work used high-performance
computing facilities operated by Center for Informatics
and Computation in Astronomy (CICA) at National Tsing Hua University (NTHU).
This equipment was funded by the MoE of Taiwan, the NSTC of 
Taiwan, and NTHU.
FLASH was developed in part by the DOE NASA- and
DOE Office of Science-supported Flash Center for Computational
Science at the University of Chicago and the
University of Rochester. 
The PRT code was developed in-house at 
    the Mullard Space Science Laboratory
    in University College London
    \citep[see][]{YLOn2019MNRAS, Chan2019MNRAS}. 
Data analysis presented in this
paper was conducted with the publicly available yt visualization
software \citep{Turk2011ApJS},
astropy (Astropy Collaboration et al. 2022)
and Google Colab. 
We are grateful to
the yt development team and community for their support.
Support for JAZ was provided by the {\it Chandra} X-ray
Observatory Center, which is operated by the Smithsonian Astrophysical Observatory for and on behalf of NASA under
contract NAS8-03060.
AYLO thanks 
    Dr. Jennifer Y. H. Chan (Dunlap/CITA),
    Paul C. W. Lai
    (MSSL/UCL),
    Prof. Kinwah Wu (MSSL/UCL),
    Dr. Ziri Younsi (MSSL/UCL),
    Kaye Li (MSSL/UCL)
    and 
    Prof. Hung-Yi Pu (NTNU)
    for 
    technical discussions
    on the PRT code
    during various stages
    of this work. 
AYLO also acknowledges
    the international
    conferences on
    Cosmic Magnetism in the Pre-SKA Era 2024
    in Kagoshima
    and
    the Cosmic Magnetism with Radio Astronomy 2024
    in Pisa
    for
    discussions
    on depolarization
    in galaxy clusters,
    in particular 
    with 
    Dr. Erik Osinga (Dunlap),
    Dr. Craig Anderson (ANU),
    Dr. Francesca Loi (INAF-OAC), 
    Dr. Yik Ki Jackie Ma (MPIfR),
    Dr. Aritra Basu (TLS)
    and
    Dr. Juan D. Soler (IAPS-INAF).
This research has made use of NASA’s Astrophysics
Data Systems.

\section*{Data Availability}

The data underlying this article will be shared on reasonable request to the corresponding authors.





%
\facilities{HST(STIS), Swift(XRT and UVOT), AAVSO, CTIO:1.3m, CTIO:1.5m, CXO}




\bibliography{references}{}

\begin{thebibliography}{}
\expandafter\ifx\csname natexlab\endcsname\relax\def\natexlab#1{#1}\fi
\providecommand{\url}[1]{\href{#1}{#1}}
\providecommand{\dodoi}[1]{doi:~\href{http://doi.org/#1}{\nolinkurl{#1}}}
\providecommand{\doeprint}[1]{\href{http://ascl.net/#1}{\nolinkurl{http://ascl.net/#1}}}
\providecommand{\doarXiv}[1]{\href{https://arxiv.org/abs/#1}{\nolinkurl{https://arxiv.org/abs/#1}}}

\bibitem[{C.~S. {Anderson} {et~al.}(2021){Anderson}, {Heald}, {Eilek}, {Lenc}, {Gaensler}, {Rudnick}, {Van Eck}, {O'Sullivan}, {Stil}, {Chippendale}, {Riseley}, {Carretti}, {West}, {Farnes}, {Harvey-Smith}, {McClure-Griffiths}, {Bock}, {Bunton}, {Koribalski}, {Tremblay}, {Voronkov}, \& {Warhurst}}]{Anderson2021PASA}
{Anderson}, C.~S., {Heald}, G.~H., {Eilek}, J.~A., {et~al.} 2021, \bibinfo{title}{{Early Science from POSSUM: Shocks, turbulence, and a massive new reservoir of ionised gas in the Fornax cluster},} \pasa, 38, e020, \dodoi{10.1017/pasa.2021.4}

\bibitem[{D.~J. {Barnes} {et~al.}(2018){Barnes}, {On}, {Wu}, \& {Kawata}}]{Barnes2018MNRAS}
{Barnes}, D.~J., {On}, A. Y.~L., {Wu}, K., \& {Kawata}, D. 2018, \bibinfo{title}{{SPMHD simulations of structure formation},} \mnras, 476, 2890, \dodoi{10.1093/mnras/sty400}

\bibitem[{L. {B{\^\i}rzan} {et~al.}(2012){B{\^\i}rzan}, {Rafferty}, {Nulsen}, {McNamara}, {R{\"o}ttgering}, {Wise}, \& {Mittal}}]{Birzan2012MNRAS}
{B{\^\i}rzan}, L., {Rafferty}, D.~A., {Nulsen}, P.~E.~J., {et~al.} 2012, \bibinfo{title}{{The duty cycle of radio-mode feedback in complete samples of clusters},} \mnras, 427, 3468, \dodoi{10.1111/j.1365-2966.2012.22083.x}

\bibitem[{A. {Bonafede} {et~al.}(2010){Bonafede}, {Feretti}, {Murgia}, {Govoni}, {Giovannini}, {Dallacasa}, {Dolag}, \& {Taylor}}]{Bonafede2010A&A}
{Bonafede}, A., {Feretti}, L., {Murgia}, M., {et~al.} 2010, \bibinfo{title}{{The Coma cluster magnetic field from Faraday rotation measures},} \aap, 513, A30, \dodoi{10.1051/0004-6361/200913696}

\bibitem[{M. {Br{\"u}ggen} {et~al.}(2012){Br{\"u}ggen}, {Bykov}, {Ryu}, \& {R{\"o}ttgering}}]{Bruggen2012SSR}
{Br{\"u}ggen}, M., {Bykov}, A., {Ryu}, D., \& {R{\"o}ttgering}, H. 2012, \bibinfo{title}{{Magnetic Fields, Relativistic Particles, and Shock Waves in Cluster Outskirts},} \ssr, 166, 187, \dodoi{10.1007/s11214-011-9785-9}

\bibitem[{B.~J. {Burn}(1966){Burn}}]{Burn1966MNRAS}
{Burn}, B.~J. 1966, \bibinfo{title}{{On the depolarization of discrete radio sources by Faraday dispersion},} \mnras, 133, 67, \dodoi{10.1093/mnras/133.1.67}

\bibitem[{C.~L. {Carilli} \& G.~B. {Taylor}(2002){Carilli} \& {Taylor}}]{Carilli2002ARA&A}
{Carilli}, C.~L., \& {Taylor}, G.~B. 2002, \bibinfo{title}{{Cluster Magnetic Fields},} \araa, 40, 319, \dodoi{10.1146/annurev.astro.40.060401.093852}

\bibitem[{J.~Y.~H. {Chan} {et~al.}(2019){Chan}, {Wu}, {On}, {Barnes}, {McEwen}, \& {Kitching}}]{Chan2019MNRAS}
{Chan}, J. Y.~H., {Wu}, K., {On}, A. Y.~L., {et~al.} 2019, \bibinfo{title}{{Covariant polarized radiative transfer on cosmological scales for investigating large-scale magnetic field structures},} \mnras, 484, 1427, \dodoi{10.1093/mnras/sty3498}

\bibitem[{T.~E. {Clarke} {et~al.}(2001){Clarke}, {Kronberg}, \& {B{\"o}hringer}}]{Clarke2001ApJ}
{Clarke}, T.~E., {Kronberg}, P.~P., \& {B{\"o}hringer}, H. 2001, \bibinfo{title}{{A New Radio-X-Ray Probe of Galaxy Cluster Magnetic Fields},} \apjl, 547, L111, \dodoi{10.1086/318896}

\bibitem[{J.~I. {Davies} {et~al.}(2013){Davies}, {Bianchi}, {Baes}, {Boselli}, {Ciesla}, {Clemens}, {Davis}, {De Looze}, {di Serego Alighieri}, {Fuller}, {Fritz}, {Hunt}, {Serra}, {Smith}, {Verstappen}, {Vlahakis}, {Xilouris}, {Bomans}, {Hughes}, {Garcia-Appadoo}, \& {Madden}}]{Davies2013MNRAS}
{Davies}, J.~I., {Bianchi}, S., {Baes}, M., {et~al.} 2013, \bibinfo{title}{{The Herschel Fornax Cluster Survey - I. The bright galaxy sample},} \mnras, 428, 834, \dodoi{10.1093/mnras/sts082}

\bibitem[{J.~W. {Dreher} {et~al.}(1987){Dreher}, {Carilli}, \& {Perley}}]{Dreher1987ApJ}
{Dreher}, J.~W., {Carilli}, C.~L., \& {Perley}, R.~A. 1987, \bibinfo{title}{{The Faraday Rotation of Cygnus A: Magnetic Fields in Cluster Gas},} \apj, 316, 611, \dodoi{10.1086/165229}

\bibitem[{M.~J. {Drinkwater} {et~al.}(2001){Drinkwater}, {Gregg}, \& {Colless}}]{Drinkwater2001ApJ}
{Drinkwater}, M.~J., {Gregg}, M.~D., \& {Colless}, M. 2001, \bibinfo{title}{{Substructure and Dynamics of the Fornax Cluster},} \apjl, 548, L139, \dodoi{10.1086/319113}

\bibitem[{A. {Dubey} {et~al.}(2009){Dubey}, {Reid}, {Weide}, {Antypas}, {Ganapathy}, {Riley}, {Sheeler}, \& {Siegal}}]{Dubey2009arXiv}
{Dubey}, A., {Reid}, L.~B., {Weide}, K., {et~al.} 2009, \bibinfo{title}{{Extensible Component Based Architecture for FLASH, A Massively Parallel, Multiphysics Simulation Code},} arXiv e-prints, arXiv:0903.4875, \dodoi{10.48550/arXiv.0903.4875}

\bibitem[{A.~S. {Eddington}(1916){Eddington}}]{Eddington1916MNRAS}
{Eddington}, A.~S. 1916, \bibinfo{title}{{The distribution of stars in globular clusters},} \mnras, 76, 572, \dodoi{10.1093/mnras/76.7.572}

\bibitem[{A.~C. {Fabian}(1994){Fabian}}]{Fabian1994ARA&A}
{Fabian}, A.~C. 1994, \bibinfo{title}{{Cooling Flows in Clusters of Galaxies},} \araa, 32, 277, \dodoi{10.1146/annurev.aa.32.090194.001425}

\bibitem[{G.~J. {Ferland} {et~al.}(2013){Ferland}, {Porter}, {van Hoof}, {Williams}, {Abel}, {Lykins}, {Shaw}, {Henney}, \& {Stancil}}]{Ferland2013RMxAA}
{Ferland}, G.~J., {Porter}, R.~L., {van Hoof}, P.~A.~M., {et~al.} 2013, \bibinfo{title}{{The 2013 Release of Cloudy},} \rmxaa, 49, 137, \dodoi{10.48550/arXiv.1302.4485}

\bibitem[{C. {Ferrari} {et~al.}(2008){Ferrari}, {Govoni}, {Schindler}, {Bykov}, \& {Rephaeli}}]{Ferrari2008ssr}
{Ferrari}, C., {Govoni}, F., {Schindler}, S., {Bykov}, A.~M., \& {Rephaeli}, Y. 2008, \bibinfo{title}{{Observations of Extended Radio Emission in Clusters},} \ssr, 134, 93, \dodoi{10.1007/s11214-008-9311-x}

\bibitem[{B. {Fryxell} {et~al.}(2000){Fryxell}, {Olson}, {Ricker}, {Timmes}, {Zingale}, {Lamb}, {MacNeice}, {Rosner}, {Truran}, \& {Tufo}}]{Fryxell2000ApJS}
{Fryxell}, B., {Olson}, K., {Ricker}, P., {et~al.} 2000, \bibinfo{title}{{FLASH: An Adaptive Mesh Hydrodynamics Code for Modeling Astrophysical Thermonuclear Flashes},} \apjs, 131, 273, \dodoi{10.1086/317361}

\bibitem[{F. {Govoni} {et~al.}(2010){Govoni}, {Dolag}, {Murgia}, {Feretti}, {Schindler}, {Giovannini}, {Boschin}, {Vacca}, \& {Bonafede}}]{Govoni2010A&A}
{Govoni}, F., {Dolag}, K., {Murgia}, M., {et~al.} 2010, \bibinfo{title}{{Rotation measures of radio sources in hot galaxy clusters},} \aap, 522, A105, \dodoi{10.1051/0004-6361/200913665}

\bibitem[{J.-H. {Ha} {et~al.}(2018){Ha}, {Ryu}, \& {Kang}}]{Ha2018ApJ}
{Ha}, J.-H., {Ryu}, D., \& {Kang}, H. 2018, \bibinfo{title}{{Properties of Merger Shocks in Merging Galaxy Clusters},} \apj, 857, 26, \dodoi{10.3847/1538-4357/aab4a2}

\bibitem[{G. {Heald}(2009){Heald}}]{Heald2009}
{Heald}, G. 2009, \bibinfo{title}{{The Faraday rotation measure synthesis technique},} in IAU Symposium, Vol. 259, Cosmic Magnetic Fields: From Planets, to Stars and Galaxies, ed. K.~G. {Strassmeier}, A.~G. {Kosovichev}, \& J.~E. {Beckman}, 591--602, \dodoi{10.1017/S1743921309031421}

\bibitem[{S. {Hutschenreuter} {et~al.}(2022){Hutschenreuter}, {Anderson}, {Betti}, {Bower}, {Brown}, {Br{\"u}ggen}, {Carretti}, {Clarke}, {Clegg}, {Costa}, {Croft}, {Van Eck}, {Gaensler}, {de Gasperin}, {Haverkorn}, {Heald}, {Hull}, {Inoue}, {Johnston-Hollitt}, {Kaczmarek}, {Law}, {Ma}, {MacMahon}, {Mao}, {Riseley}, {Roy}, {Shanahan}, {Shimwell}, {Stil}, {Sobey}, {O'Sullivan}, {Tasse}, {Vacca}, {Vernstrom}, {Williams}, {Wright}, \& {En{\ss}lin}}]{Hutschenreuter2022A&A}
{Hutschenreuter}, S., {Anderson}, C.~S., {Betti}, S., {et~al.} 2022, \bibinfo{title}{{The Galactic Faraday rotation sky 2020},} \aap, 657, A43, \dodoi{10.1051/0004-6361/202140486}

\bibitem[{L. {Iapichino} \& M. {Br{\"u}ggen}(2012){Iapichino} \& {Br{\"u}ggen}}]{Iapichino2012MNRAS}
{Iapichino}, L., \& {Br{\"u}ggen}, M. 2012, \bibinfo{title}{{Magnetic field amplification by shocks in galaxy clusters: application to radio relics},} \mnras, 423, 2781, \dodoi{10.1111/j.1365-2966.2012.21084.x}

\bibitem[{C. {Jones} {et~al.}(1997){Jones}, {Stern}, {Forman}, {Breen}, {David}, {Tucker}, \& {Franx}}]{Jones1997ApJ}
{Jones}, C., {Stern}, C., {Forman}, W., {et~al.} 1997, \bibinfo{title}{{X-Ray Emission from the Fornax Cluster},} \apj, 482, 143, \dodoi{10.1086/304104}

\bibitem[{T.~W. {Jones} \& S.~L. {O'Dell}(1977){Jones} \& {O'Dell}}]{Jones1977ApJ}
{Jones}, T.~W., \& {O'Dell}, S.~L. 1977, \bibinfo{title}{{Transfer of polarized radiation in self-absorbed synchrotron sources. I. Results for a homogeneous source.},} \apj, 214, 522, \dodoi{10.1086/155278}

\bibitem[{U. {Keshet} {et~al.}(2010){Keshet}, {Markevitch}, {Birnboim}, \& {Loeb}}]{Keshet2010ApJ}
{Keshet}, U., {Markevitch}, M., {Birnboim}, Y., \& {Loeb}, A. 2010, \bibinfo{title}{{Dynamics and Magnetization in Galaxy Cluster Cores Traced by X-ray Cold Fronts},} \apjl, 719, L74, \dodoi{10.1088/2041-8205/719/1/L74}

\bibitem[{H. {Kotarba} {et~al.}(2011){Kotarba}, {Lesch}, {Dolag}, {Naab}, {Johansson}, {Donnert}, \& {Stasyszyn}}]{Kotarba2011MNRAS}
{Kotarba}, H., {Lesch}, H., {Dolag}, K., {et~al.} 2011, \bibinfo{title}{{Galactic m{\'e}nage {\`a} trois: simulating magnetic fields in colliding galaxies},} \mnras, 415, 3189, \dodoi{10.1111/j.1365-2966.2011.18932.x}

\bibitem[{F. {Loi} {et~al.}(2025){Loi}, {Serra}, {Murgia}, {Govoni}, {Vacca}, {Maccagni}, {Kleiner}, \& {Kamphuis}}]{Loi2025AA}
{Loi}, F., {Serra}, P., {Murgia}, M., {et~al.} 2025, \bibinfo{title}{{The MeerKAT Fornax Survey: IV. A close look at the cluster physics through the densest rotation measure grid},} \aap, 694, A125, \dodoi{10.1051/0004-6361/202451711}

\bibitem[{M. {Lyutikov}(2006){Lyutikov}}]{Lyutikov2006MNRAS}
{Lyutikov}, M. 2006, \bibinfo{title}{{Magnetic draping of merging cores and radio bubbles in clusters of galaxies},} \mnras, 373, 73, \dodoi{10.1111/j.1365-2966.2006.10835.x}

\bibitem[{M. {Machacek} {et~al.}(2005){Machacek}, {Dosaj}, {Forman}, {Jones}, {Markevitch}, {Vikhlinin}, {Warmflash}, \& {Kraft}}]{Machacek2005ApJa}
{Machacek}, M., {Dosaj}, A., {Forman}, W., {et~al.} 2005, \bibinfo{title}{{Infall of the Elliptical Galaxy NGC 1404 into the Fornax Cluster},} \apj, 621, 663, \dodoi{10.1086/427548}

\bibitem[{M.~E. {Machacek} {et~al.}(2005){Machacek}, {Nulsen}, {Stirbat}, {Jones}, \& {Forman}}]{Machacek2005ApJ}
{Machacek}, M.~E., {Nulsen}, P., {Stirbat}, L., {Jones}, C., \& {Forman}, W.~R. 2005, \bibinfo{title}{{XMM-Newton Observation of an X-Ray Trail between the Spiral Galaxy NGC 6872 and the Central Elliptical Galaxy NGC 6876 in the Pavo Group},} \apj, 630, 280, \dodoi{10.1086/431944}

\bibitem[{S. {Mitton}(1971){Mitton}}]{Mitton1971MNRAS}
{Mitton}, S. 1971, \bibinfo{title}{{Observations of the distribution of polarized emission of Cygnus A at 6-cm wavelength},} \mnras, 153, 133, \dodoi{10.1093/mnras/153.2.133}

\bibitem[{M. {Murgia} {et~al.}(2004){Murgia}, {Govoni}, {Feretti}, {Giovannini}, {Dallacasa}, {Fanti}, {Taylor}, \& {Dolag}}]{Murgia2004A&A}
{Murgia}, M., {Govoni}, F., {Feretti}, L., {et~al.} 2004, \bibinfo{title}{{Magnetic fields and Faraday rotation in clusters of galaxies},} \aap, 424, 429, \dodoi{10.1051/0004-6361:20040191}

\bibitem[{J.~F. {Navarro} {et~al.}(1997){Navarro}, {Frenk}, \& {White}}]{Navarro1997ApJ}
{Navarro}, J.~F., {Frenk}, C.~S., \& {White}, S. D.~M. 1997, \bibinfo{title}{{A Universal Density Profile from Hierarchical Clustering},} \apj, 490, 493, \dodoi{10.1086/304888}

\bibitem[{A.~Y.~L. {On} {et~al.}(2025){On}, {Chan}, {Lai}, \& {Wu}}]{On2025PASA}
{On}, A. Y.~L., {Chan}, J. Y.~H., {Lai}, P. C.~W., \& {Wu}, K. 2025, \bibinfo{title}{{Depolarisation and polarisation enhancement of point sources in extended diffuse astrophysical media},} \pasa, under review

\bibitem[{A.~Y.~L. {On} {et~al.}(2019){On}, {Chan}, {Wu}, {Saxton}, \& {van Driel-Gesztelyi}}]{YLOn2019MNRAS}
{On}, A. Y.~L., {Chan}, J. Y.~H., {Wu}, K., {Saxton}, C.~J., \& {van Driel-Gesztelyi}, L. 2019, \bibinfo{title}{{Polarized radiative transfer, rotation measure fluctuations, and large-scale magnetic fields},} \mnras, 490, 1697, \dodoi{10.1093/mnras/stz2683}

\bibitem[{A.~G. {Pacholczyk}(1970){Pacholczyk}}]{Pacholczyk1970Book}
{Pacholczyk}, A.~G. 1970, Radio astrophysics: Nonthermal processes in galactic and extragalactic sources, Series of Books in Astronomy and Astrophysics (W.~H.~Freeman and Company, San Francisco), 77--111

\bibitem[{A.~G. {Pacholczyk}(1977){Pacholczyk}}]{Pacholczyk1977Book}
{Pacholczyk}, A.~G. 1977, Int. Series in Natural Philosophy, Vol.~89, Radio galaxies: Radiation transfer, dynamics, stability and evolution of a synchrotron plasmon (Pergamon Press, Oxford, New York, Toronto, Sydney, Paris, Frankfurt), 83--123

\bibitem[{E. {Roediger} \& J.~A. {Zuhone}(2012){Roediger} \& {Zuhone}}]{Roediger2012MNRAS}
{Roediger}, E., \& {Zuhone}, J.~A. 2012, \bibinfo{title}{{Fast simulations of gas sloshing and cold front formation},} \mnras, 419, 1338, \dodoi{10.1111/j.1365-2966.2011.19794.x}

\bibitem[{D. {Ryu} {et~al.}(2003){Ryu}, {Kang}, {Hallman}, \& {Jones}}]{Ryu2003ApJ}
{Ryu}, D., {Kang}, H., {Hallman}, E., \& {Jones}, T.~W. 2003, \bibinfo{title}{{Cosmological Shock Waves and Their Role in the Large-Scale Structure of the Universe},} \apj, 593, 599, \dodoi{10.1086/376723}

\bibitem[{C.~A. {Scharf} {et~al.}(2005){Scharf}, {Zurek}, \& {Bureau}}]{Scharf2005ApJ}
{Scharf}, C.~A., {Zurek}, D.~R., \& {Bureau}, M. 2005, \bibinfo{title}{{The Chandra Fornax Survey. I. The Cluster Environment},} \apj, 633, 154, \dodoi{10.1086/444531}

\bibitem[{P. {Schuecker} {et~al.}(2004){Schuecker}, {Finoguenov}, {Miniati}, {B{\"o}hringer}, \& {Briel}}]{Schuecker2004A&A}
{Schuecker}, P., {Finoguenov}, A., {Miniati}, F., {B{\"o}hringer}, H., \& {Briel}, U.~G. 2004, \bibinfo{title}{{Probing turbulence in the Coma galaxy cluster},} \aap, 426, 387, \dodoi{10.1051/0004-6361:20041039}

\bibitem[{A. {Sheardown} {et~al.}(2018){Sheardown}, {Roediger}, {Su}, {Kraft}, {Fish}, {ZuHone}, {Forman}, {Jones}, {Churazov}, \& {Nulsen}}]{Sheardown2018ApJ}
{Sheardown}, A., {Roediger}, E., {Su}, Y., {et~al.} 2018, \bibinfo{title}{{The Recent Growth History of the Fornax Cluster Derived from Simultaneous Sloshing and Gas Stripping: Simulating the Infall of NGC 1404},} \apj, 865, 118, \dodoi{10.3847/1538-4357/aadc0f}

\bibitem[{K. {Shurkin} {et~al.}(2008){Shurkin}, {Dunn}, {Gentile}, {Taylor}, \& {Allen}}]{Shurkin2008MNRAS}
{Shurkin}, K., {Dunn}, R.~J.~H., {Gentile}, G., {Taylor}, G.~B., \& {Allen}, S.~W. 2008, \bibinfo{title}{{Active galactic nuclei-induced cavities in NGC 1399 and NGC 4649},} \mnras, 383, 923, \dodoi{10.1111/j.1365-2966.2007.12651.x}

\bibitem[{M. {Simard-Normandin} \& P.~P. {Kronberg}(1980){Simard-Normandin} \& {Kronberg}}]{Simard-Normandin1980ApJ}
{Simard-Normandin}, M., \& {Kronberg}, P.~P. 1980, \bibinfo{title}{{Rotation measures and the galactic magnetic field.},} \apj, 242, 74, \dodoi{10.1086/158445}

\bibitem[{D.~D. {Sokoloff} {et~al.}(1998){Sokoloff}, {Bykov}, {Shukurov}, {Berkhuijsen}, {Beck}, \& {Poezd}}]{Sokoloff1998MNRAS}
{Sokoloff}, D.~D., {Bykov}, A.~A., {Shukurov}, A., {et~al.} 1998, \bibinfo{title}{{Depolarization and Faraday effects in galaxies},} \mnras, 299, 189, \dodoi{10.1046/j.1365-8711.1998.01782.x}

\bibitem[{Y. {Su} {et~al.}(2017{\natexlab{a}}){Su}, {Nulsen}, {Kraft}, {Forman}, {Jones}, {Irwin}, {Randall}, \& {Churazov}}]{Su2017ApJ}
{Su}, Y., {Nulsen}, P. E.~J., {Kraft}, R.~P., {et~al.} 2017{\natexlab{a}}, \bibinfo{title}{{Buoyant AGN Bubbles in the Quasi-isothermal Potential of NGC 1399},} \apj, 847, 94, \dodoi{10.3847/1538-4357/aa8954}

\bibitem[{Y. {Su} {et~al.}(2017{\natexlab{b}}){Su}, {Kraft}, {Roediger}, {Nulsen}, {Forman}, {Churazov}, {Randall}, {Jones}, \& {Machacek}}]{SuNGC14042017ApJ}
{Su}, Y., {Kraft}, R.~P., {Roediger}, E., {et~al.} 2017{\natexlab{b}}, \bibinfo{title}{{Deep Chandra Observations of NGC 1404: Cluster Plasma Physics Revealed by an Infalling Early-type Galaxy},} \apj, 834, 74, \dodoi{10.3847/1538-4357/834/1/74}

\bibitem[{Y. Su {et~al.}(2017)Su, Nulsen, Kraft, Roediger, ZuHone, Jones, Forman, Sheardown, Irwin, \& Randall}]{Su2017AAS}
Su, Y., Nulsen, P. E.~J., Kraft, R.~P., {et~al.} 2017, \bibinfo{title}{Gas Sloshing Regulates and Records the Evolution of the Fornax Cluster,} The Astrophysical Journal, 851, 69, \dodoi{10.3847/1538-4357/aa989e}

\bibitem[{S. {Sur} {et~al.}(2021){Sur}, {Basu}, \& {Subramanian}}]{Sur2021MNRAS}
{Sur}, S., {Basu}, A., \& {Subramanian}, K. 2021, \bibinfo{title}{{Properties of polarized synchrotron emission from fluctuation-dynamo action - I. Application to galaxy clusters},} \mnras, 501, 3332, \dodoi{10.1093/mnras/staa3767}

\bibitem[{G.~B. {Taylor} {et~al.}(2002){Taylor}, {Fabian}, \& {Allen}}]{Taylor2002MNRAS}
{Taylor}, G.~B., {Fabian}, A.~C., \& {Allen}, S.~W. 2002, \bibinfo{title}{{Magnetic fields in the Centaurus cluster},} \mnras, 334, 769, \dodoi{10.1046/j.1365-8711.2002.05555.x}

\bibitem[{M.~J. {Turk} {et~al.}(2011){Turk}, {Smith}, {Oishi}, {Skory}, {Skillman}, {Abel}, \& {Norman}}]{Turk2011ApJS}
{Turk}, M.~J., {Smith}, B.~D., {Oishi}, J.~S., {et~al.} 2011, \bibinfo{title}{{yt: A Multi-code Analysis Toolkit for Astrophysical Simulation Data},} \apjs, 192, 9, \dodoi{10.1088/0067-0049/192/1/9}

\bibitem[{R.~J. {van Weeren} {et~al.}(2010){van Weeren}, {R{\"o}ttgering}, {Br{\"u}ggen}, \& {Hoeft}}]{van2010Sci}
{van Weeren}, R.~J., {R{\"o}ttgering}, H. J.~A., {Br{\"u}ggen}, M., \& {Hoeft}, M. 2010, \bibinfo{title}{{Particle Acceleration on Megaparsec Scales in a Merging Galaxy Cluster},} Science, 330, 347, \dodoi{10.1126/science.1194293}

\bibitem[{L.~M. {Widrow}(2002){Widrow}}]{Widrow2002Reviews}
{Widrow}, L.~M. 2002, \bibinfo{title}{{Origin of galactic and extragalactic magnetic fields},} Reviews of Modern Physics, 74, 775, \dodoi{10.1103/RevModPhys.74.775}

\bibitem[{J.~A. {ZuHone} {et~al.}(2011){ZuHone}, {Markevitch}, \& {Lee}}]{ZuHone2011ApJ}
{ZuHone}, J.~A., {Markevitch}, M., \& {Lee}, D. 2011, \bibinfo{title}{{Sloshing of the Magnetized Cool Gas in the Cores of Galaxy Clusters},} \apj, 743, 16, \dodoi{10.1088/0004-637X/743/1/16}

\bibitem[{J.~A. {ZuHone} {et~al.}(2013){ZuHone}, {Markevitch}, {Ruszkowski}, \& {Lee}}]{ZuHone2013ApJ}
{ZuHone}, J.~A., {Markevitch}, M., {Ruszkowski}, M., \& {Lee}, D. 2013, \bibinfo{title}{{Cold Fronts and Gas Sloshing in Galaxy Clusters with Anisotropic Thermal Conduction},} \apj, 762, 69, \dodoi{10.1088/0004-637X/762/2/69}

\end{thebibliography}
\bibliographystyle{aasjournalv7}



\end{document}